\documentclass{article}

\usepackage{arxiv}

\usepackage[utf8]{inputenc} 
\usepackage[T1]{fontenc}    
\usepackage{hyperref}       
\usepackage{url}            
\usepackage{booktabs}       
\usepackage{amsfonts}       
\usepackage{nicefrac}       
\usepackage{microtype}      
\usepackage{lipsum}		
\usepackage{enumitem}

\usepackage{amsmath}
\usepackage{graphics}
\usepackage{times}
\usepackage{bm}
\usepackage{natbib}
\usepackage{booktabs}
\usepackage{multirow}
\usepackage{color}
\usepackage{url}
\usepackage[demo]{rotating}

\graphicspath{{./art/}}

\usepackage[plain,noend]{algorithm2e}

\newtheorem{condition}{Condition}
\newtheorem{remark}{Remark}
\newtheorem{theorem}{Theorem}
\newtheorem{lemma}{Lemma}

\newcommand{\lnorm}{\bigg\|}
\usepackage{latexsym}
\usepackage{amsmath}
\usepackage{amssymb}
\usepackage{amsfonts}
\usepackage{bm}

\DeclareMathOperator*{\argmax}{arg\,max}

\newcommand{\indep}{\perp \!\!\! \perp}

\newcommand{\lp}{\left(}
\newcommand{\rp}{\right)}
\newcommand{\llp}{\left\{}
\newcommand{\rrp}{\right\}}
\newcommand{\lllp}{\left[}
\newcommand{\rrrp}{\right]}
\newcommand{\pd}{\partial}
\newcommand{\eff}{\mathrm{eff}}

\newcommand{\KH}{\mathrm{KH}}

\newcommand{\EL}{\mathrm{EL}}
\newcommand{\MM}{\mathrm{MM}}

\newcounter{defn}
\renewcommand{\thedefn}{\arabic{defn}}

\newenvironment{defn}[1][Definition]{%
  \refstepcounter{defn}
  \noindent\textbf{#1~\thedefn.} \normalfont%
}{\par}

\renewenvironment{remark}[1][Remark]{%
  \begin{trivlist}%
  \item[\hskip \labelsep {\bfseries\itshape #1}]%
}{%
  \end{trivlist}%
}

\title{Efficient Multiple-Robust Estimation for Nonresponse Data Under Informative Sampling}

\author{%
  \href{https://orcid.org/0000-0002-6021-5180}{\includegraphics[scale=0.06]{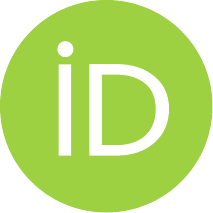}\hspace{1mm}Kosuke Morikawa}\textsuperscript{1,2}
  \hspace{2mm}Kenji Beppu\textsuperscript{2}
  \hspace{2mm}and\hspace{1mm}Wataru Aida\textsuperscript{2}\\[2mm]
  \textsuperscript{1} Department of Statistics, Iowa State University, Ames, IA 50010, USA\\
  \textsuperscript{2} Graduate School of Engineering Science, the University of Osaka, Toyonaka, Osaka 560-8531, Japan\\
  \texttt{morikawa@iastate.edu}
}

\renewcommand{\shorttitle}{Efficient estimation for nonresponse data under informative sampling}

\hypersetup{
pdftitle={Efficient estimation for nonresponse data under informative sampling},
pdfsubject={q-bio.NC, q-bio.QM},
pdfauthor={David S.~Hippocampus, Elias D.~Striatum},
pdfkeywords={First keyword, Second keyword, More},
}

\begin{document}
\maketitle

\begin{abstract}
	Nonresponse in probability sampling presents a long-standing challenge in survey sampling, often necessitating simultaneous adjustments to address sampling and selection biases. We develop a statistical framework that explicitly models sampling weights as random variables and establish the semiparametric efficiency bound for the parameter of interest under nonresponse. This study investigates strategies for eliminating bias and effectively utilizing available information, extending beyond nonresponse issues to data integration with external summary statistics. The proposed estimators are characterized by their efficiency and double robustness. However, realizing full efficiency hinges on the accurate specification of underlying models. To enhance robustness against potential model misspecification, we expand double robustness to multiple robustness through a novel two-step empirical likelihood approach. A numerical study evaluates the finite-sample performance of our methods. Additionally, we apply these methods to a dataset from the National Health and Nutrition Examination Survey, effectively integrating summary statistics from the National Health Interview Survey.
\end{abstract}


\section{Introduction}

    Survey sampling is a foundational pillar of modern statistics and the social sciences, providing crucial information without the need to conduct a complete census \citep[Chap. 1]{Chambers2003, fuller09}.
    While the history of survey sampling is deeply rooted in the field of statistics, its development continues to evolve.
    Ongoing adjustments in sampling methodologies are essential to account for changes in privacy norms and other societal dynamics. These shifts have heightened the complexity and significance of addressing nonresponse.

    Probability sampling, accompanied by sampling weights derived as the inverse of the selection probabilities, is the most commonly applied method in practice. These weights are typically determined based on available covariates prior to observing outcome variables. When sampling weights are independent of the outcomes given the available covariates, the sampling mechanism is referred to as non-informative. However, if analysts do not have access to some of these covariates, the weights may become dependent on the outcomes, resulting in informative sampling \citep{Pfeffermann1993, pferffermann09}. This dependence can distort the distribution of outcome variables among sampled units compared to the target population, potentially leading to biased estimates. Previous studies, including \citet{HT1952}, \citet{Binder1983}, and \citet[p. 236]{hajek71}, have effectively addressed this selection bias by appropriately adjusting the sampling weights.

However, adjusting sampling weights alone is insufficient to counteract biases resulting from nonresponse. It is crucial to accurately model the response mechanism to understand why an outcome variable is observed; a misspecified model can lead to biased estimates. The double-robust estimator proposed by \citet{kim14_dr} ensures consistency provided that either the response or outcome regression model is correctly specified, thus mitigating the risk of misspecification. This methodology was further refined by \citet{chen17_mr}, who introduced a multiple-robust estimator capable of handling multiple model candidates for both the response and outcome regression models.

In terms of efficiency, \citet{godambe09} developed an optimal estimator within a class of design-unbiased estimating functions, assuming non-random sampling weights. Subsequent advancements by \citet{pfe99}, \citet{pferffermann09}, and \citet{kim13} have yielded more efficient estimators through smoothing the sampling weights. Our study builds upon the semiparametric framework for informative sampling developed by \citep{morikawa22}, which assumes a Poisson sampling mechanism and independent and identically distributed observations from the superpopulation model \citep{robins94,rotnitzky1997analysis}. Under these same assumptions, we introduce semiparametric efficient estimators specifically designed for scenarios involving nonresponse. We present adaptive estimators that achieve the semiparametric lower bound across various settings, as illustrated in Figure \ref{fig:1}. Although our approach requires specifying both response and outcome regression models, we employ multiple-robust estimators to effectively mitigate potential biases arising from model misspecification.

Furthermore, recent developments in data integration by \citet{chatterjee2016}, \citet{kundu19}, and \citet{zhang2020generalized} have combined individual internal data with summary statistics from external sources, such as means and variances of covariates and outcome variables. Drawing on the results of Hu et al. (2022), who identified the efficiency bound in data integration settings, we derive optimal estimators under informative sampling in cases where external summary statistics are available.

\section{Basic Setup}
\subsection{Notation}

Suppose that there exists a superpopulation of random variables $ (X, Y, Z) $, where $ Y $ denotes an outcome, and $ X $ and $ Z $ are explanatory variables. We draw  identically and independently distributed $N$ copies $\{X_i, Y_i, Z_i\}_{i=1}^N $ from the superpopulation. Our aim is to estimate a parameter $ \theta $ that characterizes the relationship between $ X $ and $ Y $. The target parameter $\theta^*$ is uniquely determined by the solution to the equation $ E\{U_\theta(X, Y)\} = 0 $. For example, if our focus is on $ E(Y) $, then $ U_\theta(y) = y - \theta $, and if our interest lies in the regression parameter $ \theta $ within $ \mu(x; \theta) = E(Y \mid x; \theta) $, then $ U_\theta(x,y) = A(x)\{y - \mu(x; \theta)\} $, where $ A(x) $ is a function of $ x $ with the same dimensionality as $ \theta $.
Note that we allow $X$ and $Z$ to be continuous or discrete, and possibly multivariate. While the variable $Y$ is continuous or discrete, we restrict attention to the univariate case to avoid the increased complexity of missingness patterns in multivariate outcomes.

We define $\delta$ as a sampling indicator, which is equal to one if a unit is sampled and zero otherwise. In probability sampling, we draw a sample of size $n$ $(< N)$ from a finite population based on inclusion probabilities given by $1/W$, where $W$ denotes sampling weights usually constructed using covariates $X$ and $Z$ obtained in advance. Then, we assume that the selection probability is determined by the sampling weight: $P(\delta=1 \mid X,Y,Z,W)=P(\delta=1 \mid W)=1/W$.
Throughout this study, we assume a Poisson sampling mechanism, i.e.,  the indicators $\delta_i$ are independent conditional on the sampling weights. It might seem counterintuitive to treat $W$ as a random variable; however, this perspective arises naturally. We begin by defining informative sampling and then illustrate how this formulation emerges from that setting.

\begin{defn} \label{d1}
Sampling mechanism is \textit{informative} if, in Setting 1, $ W \not \! \indep (Y, Z)\mid X $, and in Settings 2 and 3, $ W \not\!\indep Y \mid (X, Z)$ hold.
\end{defn}

When the negation of definition \ref{d1} holds in Setting 1, we have
$$
P(\delta=1\mid x,y,z) = E(W^{-1}\mid x, y, z)=E(W^{-1}\mid x)=P(\delta = 1\mid x).
$$
This sampling mechanism is referred to as non-informative sampling in the literature of survey sampling \citep{Pfeffermann1993} and as missing at random in missing data analysis \citep{rubin1976}. Under this sampling mechanism, we have $f(y\mid x, \delta = 1) = f(y\mid x)$.
Therefore, the sampled outcome distribution is the same as that of the population. In contrast, definition \ref{d1} allows the two distributions to differ, and ignoring the sampling mechanism may result in biased results. 

\begin{figure}[htbp]
		 \begin{center}
		  \includegraphics[width=100mm]{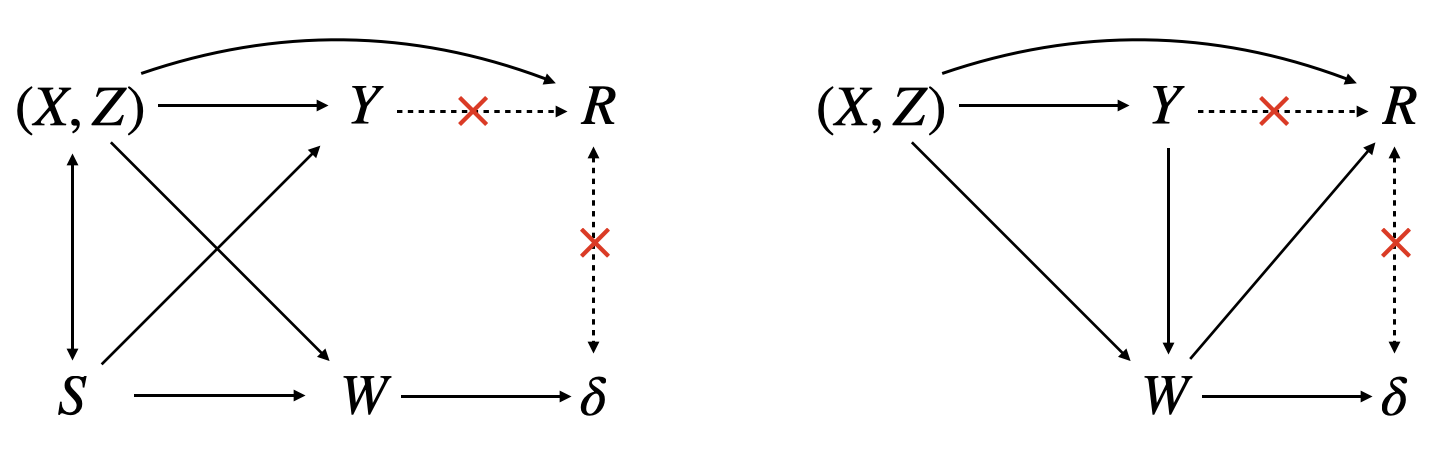}
		 \end{center}
		 \caption{Directed acyclic graphs when it is appropriate to treat $W$ as a random variable. On the left is the perspective of the sampling weight designer, and on the right is the perspective of the data analyst.
         The dashed lines (crossed out in red) represent dependence relations excluded under the PMAR and RSCI assumptions, while solid arrows denote direct dependence.}
    \label{dag}
\end{figure}

Here, we illustrate a situation in which Definition \ref{d1} applies. Let $S$ denote a covariate available to a sampling designer who constructs the sampling weights.
The left side of Figure \ref{dag} shows the directed acyclic graph (DAG) representing the variables accessible to the sampling designer.
In this scenario, the data analyst is considered distinct from the sampling designer and does not have access to the variable $S$; this perspective is reflected in the DAG on the right side of Figure \ref{dag}.
The sampling design is informative because the sampling weights $W$ remain associated with the study outcome even after adjusting for covariates.
As a result, $W$ becomes correlated with the outcome and, from the analyst’s perspective, is best treated as a random variable—since the analyst does not know how $W$ was constructed.
Unlike in classical design-based inference, where design weights are treated as fixed, this approach recognizes the influence of unobserved variables in the sampling process.

\begin{figure}[htbp]
		 \begin{center}
		  \includegraphics[width=112mm]{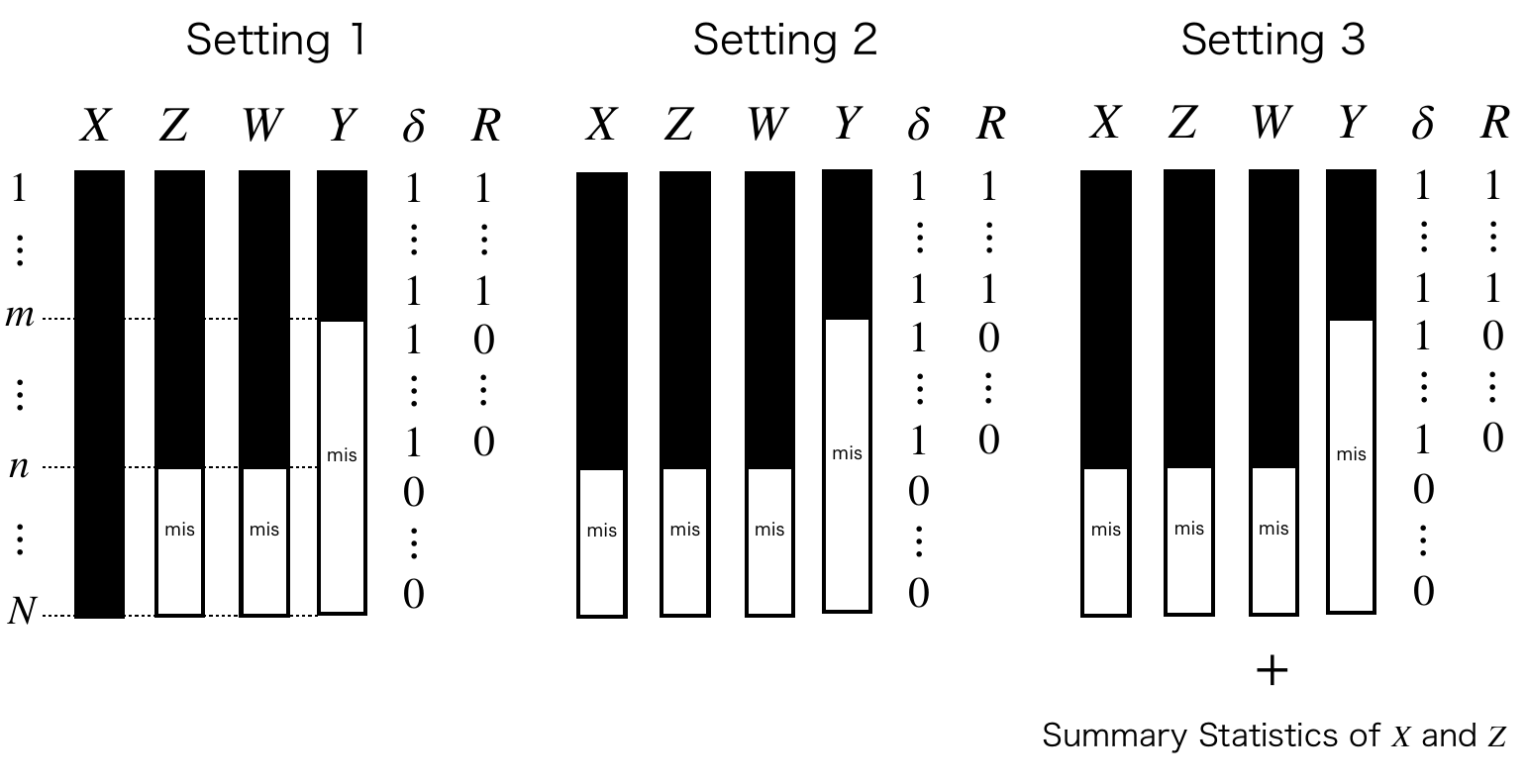}
		 \end{center}
		 \caption{Three settings considered in this study. The data highlighted in black represents observed data, and the entries labeled as ``mis" indicate unsampled or nonresponse.}
    \label{fig:1}
\end{figure}

Consider a scenario in which the outcome variable $Y$ is subject to missingness among the sampled units, resulting in only $ m $ out of $ n $ units having fully observed data. Let $ R $ be a response indicator for $ Y $, which is equal to one when $ Y $ is observed and equal to zero otherwise.
We outline our setup across three distinct settings in Figure \ref{fig:1}. In Setting 1, information regarding $ X $ is available for all units. In Setting 2, information regarding $ X $ is limited to sampled units. In Setting 3, instead of using data from $ X $ for unsampled units, we utilize additional data sources that provide information such as the mean and variance of $ X $.
For notational simplicity, without loss of generality, we rearrange the order of the units as follows: the initial $ m $ units are fully observed, the subsequent $ n - m $ units are sampled with $ Y $ units missing, and the remaining $ N - n $ units are unsampled. 

We denote the independence of two random variables $X$ and $Y$ as $X\indep Y$, and postulate the following assumption regarding the distribution of the superpopulation of $ (X, Y, Z, W, \delta, R) $:

\begin{condition}[Population missing at random]
\label{c1}
$ R \indep Y \mid (X, Z, W)$.
\end{condition}
\begin{condition}[Response-sampling conditional independence]
\label{c2}
$R\indep \delta \mid (X, Z, Y, W)$.
\end{condition}

Condition~\ref{c1} corresponds to the condition $R\indep Y \mid (X,Z)$ defined in \citet*{berg16}, extended by conditioning further on $W$.  
Although it may appear unconventional to include the sampling weight $W$ in the conditioning sets of PMAR, in the presence of missing data under informative sampling, there are additional advantages to conditioning on $W$. As illustrated in the left panel of Figure~1, beyond the previously noted indirect dependence between $Y$ and $W$, the unobservability of the design variable $S$ also induces an indirect dependence between $W$ and $R$, because $S$ connects them through the paths $S \to W$ and $S \to (X,Z) \to R$.  
Therefore, even if $Y\indep R\mid (X,Z,S)$ holds, $Y\indep R\mid (X,Z)$ does not necessarily hold.  
Our PMAR condition is formulated to appropriately eliminate the influence of such unobserved variables $S$, ensuring valid inference.

Condition~\ref{c2}, the response–sampling conditional independence (RSCI), assumes that conditional on a unit’s background characteristics, whether the unit is sampled and whether it responds are independent.  
This is reasonable because sampling is determined by the survey design, whereas responding is driven by individual behavior—two conceptually distinct mechanisms.  
This additional condition bridges the gap between PMAR and the sample missing-at-random (SMAR) condition, $R \indep Y \mid (X,Z,W,\delta=1)$, which constrains the response mechanism of subsequent missingness to be missing at random \citep{little03, Pfeffermann1993}.

The directed acyclic graph (DAG) in the right panel of Figure~\ref{dag} illustrates the implied conditional independences.  
As a theoretical implication, we note that RSCI and PMAR together imply $R \indep (Y,\delta)\mid (X,Z,W)$, which in turn allows the response model to be specified conditional only on $(X,Z,W)$, without reference to the unobserved outcome $Y$ or $\delta$; that is, $P(R=1\mid X,Z,W)$.

\subsection{Estimators with nonresponse under informative sampling}
Under Conditions \ref{c1} and \ref{c2}, if the response mechanism $\pi(x,z,w)=P(R=1\mid x,z,w)$ is known, the parameter $\theta$ can be estimated by using the inverse probability weighted estimating equation \citep{Binder1983}:
\begin{align}
\sum_{i=1}^N \frac{\delta_i W_i R_i}{\pi(X_i, Z_i, W_i)} U_\theta(X_i, Y_i)
=\sum_{i=1}^{m}   \frac{W_i}{\pi(X_i, Z_i, W_i)} U_\theta(X_i, Y_i)
=0.\label{HT}
\end{align}
It is worth noting that the response mechanism is often defined through covariates $X$ and $Z$, independent of sampling weights. We will show later that incorporating sampling weights into the modeling of the response mechanism yields a more efficient estimator.

However, the response mechanism is generally unknown and must be modeled and estimated. The parameters of the response model can be estimated using maximum likelihood estimation, as covariates $X$, $Z$, and $W$ are observed for all units where $\delta = 1$. Given that misspecification of the response model can lead to biased results, we initially consider a double-robust estimator:
\begin{align}
\sum_{i=1}^N \delta_i W_i 
\biggl[
\frac{R_i}{\hat{\pi}(X_i, Z_i, W_i)} U_\theta(X_i, Y_i) 
+ 
\biggl\{
1-\frac{R_i}{\hat{\pi}(X_i, Z_i, W_i)}
\biggr\}
\hat{g}_\theta(X_i, Z_i, W_i) 
\biggr]
=0, \label{KH}
\end{align}
where $g_\theta(x,z,w) = E\{U_\theta(x,Y)\mid x,z,w\}$ and  ``hat" denote estimated parametric models based on observed data (e.g., $\hat{\pi}(x,z,w)$ and $\hat{g}_\theta(x,z,w)$ are estimated functions using $\pi(x,z,w)$ and $g_\theta(x,z,w)$, where the detailed procedure is discussed in \S \ref{sec:4}). \citet{kim14_dr} considered a similar estimator without including weights in the response and the regression models.
Although nonparametric working models are also usable due to orthogonality to the nuisance parameter \citep{chernozhukov18}, we focus on parametric working models in this paper. The solution $\hat{\theta}_{\KH}$ to the estimating equation \eqref{KH} has double robustness: if either $\pi(x,z,w)$ or $g_\theta(x,z,w)$ is correctly specified, then it has consistency. However, this estimator is not necessarily efficient, even if both models are correct, unlike the ordinal double-robust estimator \citep*{robins94}, because it does not leverage the information in the data of $\delta_i=0$ (i.e., $X_i\,(i=n+1, \dots, N)$) in Setting 1 or the information of $N$ in Settings 2 and 3.  

Hereafter, we derive the semiparametric efficiency bound of the parameter $\theta$ in Settings 1--3 in Figure \ref{fig:1} and propose adaptive estimators that can attain the lower bound by using the method of moments and empirical likelihood.

\section{Semiparametric Efficiency Bounds}
\label{sec:3}
\subsection{Efficiency bounds in Settings 1 and 2}
We confine our focus to Setting 1 in this section, and for Setting 2, only theoretical results are presented because the derivation is analogous. The semiparametric efficiency bound is the smallest asymptotic variance of estimators in a class of regular and asymptotically linear estimators \citep[Chap. 1]{bickel1998, tsiatis2006}. A key concept in deriving the lower bound in our study is to consider the survey weight as a random variable. \citet*{morikawa22} derived the semiparametric efficiency bound in the absence of nonresponse leveraging the general framework of \cite{robins94} and \cite{rotnitzky1997analysis}.
In this case, the optimal estimator is obtained by solving
\begin{align}
\sum_{i=1}^N \llp\delta_i W_i  U_\theta (X_i, Y_i) + (1-\delta_i W_i)C_\theta(X_i) \rrp = 0,\label{morikawa}
\end{align}
where $C_\theta(x)=E\{(W-1)U_\theta(x,Y)\mid x\}/E(W-1\mid x)$. The first term is the Horvitz-Thompson estimator, and the second term efficiently incorporates the information on unsampled $X$ into the estimating equation. It is crucial to mention that the second term always has a mean of zero, so it is termed ``augmented term". Intuitively, when $W$ is correlated with $Y$, it is expected that incorporating the information on $W$ into the model yields a more efficient estimator.

Let $f(w, z\mid x, y; \eta_1)$ and $f(y, x; \eta_2, \theta)$ be the density functions of $[(w, z)\mid (x,y)]$ and $[y, x]$, respectively, where $\eta_1$ and $\eta_2$ are infinite-dimensional nuisance parameters. The model for the superpopulation is
$
 f(x,y,z,w; \theta, \eta_1, \eta_2)
 = f(z, w\mid x,y; \eta_1)f(y, x; \theta, \eta_2)$.
whereas the model for the observed data under Setting 1 in Figure \ref{fig:1} is
\begin{align*}
L(\theta,\eta)
&=\llp \frac{1}{w}\pi(x,z,w;\psi)f(x,y,z,w; \theta, \eta_1, \eta_2)\rrp^{\delta r} \\
&\quad \times 
\biggl[
\frac{1}{w}
\{
1-\pi(x,z,w;\psi)
\}
f(x,z,w; \theta, \eta_1, \eta_2)
\biggr]
^{\delta(1-r)}\\
&\quad \times 
\biggl[
\int\int \lp 1-\frac{1}{w} 
\rp
\{
1-\pi(x,z,w;\psi)
\}
f(x,z,w; \theta, \eta_1, \eta_2)\mathrm{d}z\mathrm{d}w
\biggr]
^{1-\delta}. \end{align*}
where $\pi(x,z,w;\psi)=P(R=1\mid x,z,w,\delta=1;\psi)$ is a parametric model that is independent of $Y$ according to Conditions \ref{c1} and \ref{c2}. The semiparametric model for observed data under Setting 2 is obtained similarly by integrating the function for $\delta = 0$ with respect to $x$.

We derive the efficiency bounds for $\theta$ not unaffected by the nuisance parameters. Let each $\Lambda_\eta$ and $\Lambda_\psi$ be the nuisance tangent spaces for $\eta$ and $\psi$, respectively. Then, the nuisance tangent space $\Lambda$ is represented by the direct sum $\Lambda = \Lambda_\eta \oplus \Lambda_\psi$. By letting the score function be $S_\theta=\pd \log L(\theta,\eta)/\pd \theta$, we can obtain an efficient score function by projecting $S_\theta$ onto the orthogonal space of $\Lambda$. Let $q$ be the dimension of the parameter $\theta$ of interest and $\mathcal{H}$ be a Hilbert space that comprises the $q$-dimensional measurable functions $h=h(X, Y, Z, W, \delta, R)$ such that $E(h)=0$ and $\langle h, h\rangle <\infty$, where the inner product is defined by $\langle h, l\rangle := E(h^\top l)$ for any $h,l\in\mathcal{H}$. Furthermore, let $\Lambda^{sp,\perp}$ and $\Lambda^\perp$ be the orthogonal nuisance tangent spaces for the superpopulation and the sampled data, respectively. The projection onto a subspace $\tilde{\mathcal{H}} \subset \mathcal{H}$ is denoted by $\Pi(\cdot\mid \tilde{\mathcal{H}})$. Let the efficient score in the superpopulation model be $S^{sp}_{\eff,\theta}=\prod(S_{\theta}\mid \Lambda^{sp,\perp})$ and that in the observed model, which is our focus here, be $S_{\eff,\theta}=\prod(S_{\theta}\mid \Lambda^\perp)$.

\begin{theorem}
\label{thm.3.1}
The efficient score function in Setting 1 is
\begin{align}
S_{\eff,\theta}(\delta, R, X, Y, Z, W)= \delta W D_\theta(R, X, Y, Z, W)  + (1-\delta W) C_\theta(X), \label{eff_M}
\end{align}
where $C_\theta(x)$ is already defined in \eqref{morikawa} and
$D_\theta(r, x, y, z, w) = rU_\theta(x, y)/\pi(x, z, w) + \{1 -  r/\pi(x, z, w)\}g_\theta(x,z,w)$.
The efficient score function in Setting 2 is \eqref{eff_M} with the same $D_\theta$ as above but different $C_\theta(x)=C_\theta=E\{(W-1)U_\theta(X, Y)\}/E(W-1)$. The efficient influence function is
$\varphi_{\eff,\theta}=B_\theta S_{\eff,\theta}$, and the semiparametric efficiency bound for $\theta$ is $\{E(\varphi_{\eff,\theta^*}^{\otimes 2})\}^{-1}$, where $B_\theta=E\{\pd U_\theta(X,Y)/\pd \theta^\top\}^{-1}$ and $B^{\otimes 2}=BB^\top$ for any matrix $B$.
\end{theorem}

The efficient score is endowed with several intriguing characteristics. Most importantly, the precision of \( C_\theta(X) \) or \( C_\theta \) has a pronounced effect on efficiency without compromising consistency, as underscored by \citet*{morikawa22}. When both \( R \equiv 1 \) and \( \pi(X,Z,W) \equiv 1 \) hold, \( D_\theta(R,X,Y,Z,W) = U_\theta(X, Y) \), which gives rise to the optimal estimating equation in \citet*{morikawa22}. Setting \( C_\theta(X) = 0 \) still produces a double-robust estimator, as delineated in \eqref{KH}. This characteristic highlights the dual benefits of the efficient score, namely its optimality and the preservation of double robustness. Implementing sampling weights can enhance estimator efficiency and mitigate bias in the presence of nonresponse, particularly when Conditions \ref{c1} and \ref{c2} are violated by the dependency of design variables, which are unavailable for data analysts, on the outcome variable.

\subsection{Efficiency bound in Setting 3}

In Setting 3, to make summary statistics from additional data sources available, we must extend the class of estimators. Let $P_0 \in \mathcal{P}_0$ and $P_1 \in \mathcal{P}_1$ be the internal and external distributions, respectively, where $\mathcal{P}_0$ and $\mathcal{P}_1$ are collections of probability distributions.
We define the summary statistics $\tau^*=\tau(P_1)$ in the external data source. Additionally, we denote the estimator with an external data source of sample size $N_1$ as $\tilde{\tau}=\tilde{\tau}(P_1)$, which emphasizes that the parameter is estimated from the external distribution $P_1$. Integrating multiple data sources is also possible, but we confine our analysis to a single data source for the sake of simplicity. Readers can consult the discussion in \citet{hu2022paradoxes} for additional details on the extension to multiple data sources.

Following \citet{hu2022paradoxes}, we adopt three assumptions regarding the summary statistics from the external data source.
\begin{condition}
    \label{c3}
   The summary statistics $\tau=\tau(P_1)$ represent a parameter of the fully observed covariates $X$ and $Z$;
\end{condition}
\begin{condition}
  The estimated summary statistics $\tilde{\tau}$ represent a regular and asymptotically linear estimator of $\tau(P_1)$, and $N_1^{1/2} \{ \tilde{\tau} - \tau(P_1) \}$ converges weakly to the normal distribution with mean zero and variance $\Sigma_1$, where $\Sigma_1$ is the asymptotic variance of $\tilde{\tau}$ and a consistent estimator $\tilde{\Sigma}_1$ for $\Sigma_1$ is available;\label{c4}
\end{condition}
\begin{condition}
  The sample size from the external data source $N_1=N_{1,N}$ satisfies $N_{1,N}/N\to \rho\in(0,\infty)$ as $N\to\infty$;\label{c5}
\end{condition}
\begin{condition}
  $\tau(P_0)=\tau(P_1)$. \label{c6}
\end{condition}
Condition \ref{c3} confines the summary statistics to the sampled data without missingness, such as sample means of $X$ and $Z$, and regression coefficients $Z$ on $X$, and excludes statistics related to outcome variables.
Condition \ref{c4} is a regular condition for estimators. One may feel that Condition \ref{c5} is strange because internal and external data are often independent, but this assumption requires the sample size to go to infinity according to the ratio $\rho$. This condition is necessary to investigate the large-sample property and reflect the difference between the sample sizes $N_1$ and $N$.
Condition \ref{c6} requires the consistency of the target parameter between two data sources. Specifically, the populations can differ, but the target parameters must be the same.

Let $I_i =(X_i, Y_i, Z_i, W_i, \delta_i, R_i)\,(i=1,\dots, N)$ and  $E_i\, (i=1,\dots, N_1)$ be random vectors in the internal and the external dataset, respectively. Then, our estimator for $\theta$ can be represented as $\hat{\theta}_N=\hat{\theta}_N(I_1, \dots, I_N, \tilde{\tau})$ because it depends on both the internal data $\{I_i\}_{i=1}^N$ and summary statistics $\tilde{\tau}$ from the external data. We assume that our estimator in Setting 3 is in the class of data-fused regular and asymptotically linear estimators \citep{hu2022paradoxes}.\\

\begin{defn}
\textit{[Data-Fused Regular and Asymptotically Linear Estimator]}
An estimator $\hat{\theta}_{N}=\hat{\theta}_{N}(I_1, \dots, I_N, \tilde{\tau})$ is said to be \textit{data-fused regular and asymptotically linear} if the following two conditions hold:
    \begin{itemize}
        \item[(i)] (\textit{Regular}).
        Let $\xi$ be a finite-dimensional parameter in any parametric sub-model $P_0(I;\xi )\times P_1(E; \xi)\in \mathcal{P}_0\times \mathcal{P}_1$ and $\xi^*$ be the true value. Then, 
        $N^{1/2} \{\hat{\theta}_N( I_1^{(N)}, \ldots, I_N^{(N)}, \tilde{\tau}^{(N_1)}  ) - \theta( P_0(I;\xi_n) ) \}$
        has a limiting distribution that does not depend on the local data generation process, in which the data $\{ I_1^{(N)}, \ldots, I_N^{(N)} \}$ and $\{ E_1^{(N_1)}, \ldots, E_{N_1}^{(N_1)} \}$ are i.i.d. samples from $P_0(I; \xi_n)$ and $P_1(E; \xi_n)$, $\tilde{\tau}^{(N_1)}$ denotes the summary statistics estimated from $\{E_1^{(N_1)}, \ldots, E_{N_1}^{(N_1)} \}$, $N_1/N\to \rho$, and $N^{1/2}(\xi_n - \xi^* )$ converges to a constant.
        \item[(ii)] (\textit{Asymptotically linear}). The estimator $\hat{\theta}_N$ has the form
        $\hat{\theta}_N= \theta^* + N^{-1}\sum_{i=1}^{N} \psi_0(I_i) + \psi_1(\tilde{\tau}) + o_p(N^{-1/2}),$
        where $E\{ \psi_0(I) \}=0$, $E\{ \psi_0(I)^{\otimes 2} \}$ is finite and non-singular, $\psi_1(\tilde{\tau})$ is continuously differentiable, and $\psi_1(\tau(P_1))=0$.
    \end{itemize}
\end{defn}

\citet{hu2022paradoxes} extended the convolution theorem for the ordinary class of regular and asymptotically linear estimators to data-fused regular and asymptotically linear estimators and derived the efficiency bound of the new class. 

\begin{lemma}
    Suppose $\hat{\theta}_N$ is a data-fused regular and asymptotically linear estimator and Conditions \ref{c3}-\ref{c6} hold. Then, we have
    \begin{align*}
        N^{1/2} \begin{pmatrix}
            \hat{\theta}_N - \theta^* -N^{-1} \sum_{i=1}^{N} (\phi_{\eff,i} - M\eta_{\eff,i}) - M(\tilde{\tau} - \tau) \\
            N^{-1}\sum_{i=1}^{N} (\phi_{\eff,i} - M\eta_{\eff,i}) +M(\tilde{\tau} - \tau)
        \end{pmatrix}
        \to
        \begin{pmatrix}
            \Delta_0 \\ \Delta_1
        \end{pmatrix}
    \end{align*}
    in distribution, where
 $M=E(\phi_{\eff} \eta_{\eff}^{\top}) \{  \Sigma_1/\rho +E(\eta_{\eff}^{\otimes 2}) \}^{-1}$, $\Delta_0$ and $\Delta_1$ are independent random variables, and $\phi_{\eff,i}$ and $\eta_{\eff,i}\,(i=1,\dots, N)$ are efficient influence functions for $\theta$ and $\tau$ based on the internal data.
\end{lemma}
The convolution theorem yields the semiparametric efficiency bound in Setting 3 as $
E(\phi_{\eff}^{\otimes 2} ) - E(\phi_{\eff} \eta_{\eff}^{\top})\{ \Sigma_1/\rho + E(\eta_{\eff}^{\otimes 2}) \}^{-1} E(\phi_{\eff} \eta_{\eff}^{\top})^{\top}$
in the class of data-fused regular and asymptotically linear estimators.
The first term is the efficiency bound when using only internal data. This implies that incorporating external summary statistics into estimation always results in a more efficient estimator. 
Also, the convolution theorem in the data fusion setting reveals how the internal and external data sources contribute jointly to the efficiency of the estimator. The efficiency bound is determined not only by the internal influence function $\phi_{\text{eff}}$ but also by the covariance with the external influence function $\eta_{\text{eff}}$. Specifically, if $\phi_{\text{eff}}$ and $\eta_{\text{eff}}$ are strongly correlated, the external summary statistics substantially reduce the overall variance. Conversely, when this dependence is weak, the efficiency improvement is limited. This highlights the importance of leveraging external information that is complementary to the internal data in terms of its inferential value.

\section{Adaptive Estimator}
\label{sec:4}
We derive adaptive estimators under Settings 1--3 that can attain the semiparametric efficiency bounds derived in \S \ref{sec:3}. To construct adaptive estimators, we require some working models for $\pi(x,z,w;\alpha)$, $g_\theta(x,z,w;\beta)$, and $C_\theta(X;\gamma)$, where $\alpha$, $\beta$, and $\gamma$ are unknown finite-dimensional parameters. Each model is naturally estimated based on the ignorability (Conditions \ref{c1} and \ref{c2}): $\hat{\alpha}$ is the maximum likelihood estimator,
$$\hat{\alpha} = \argmax_{\alpha} \prod_{i=1}^n \pi(X_i, Z_i, W_i;\alpha)^{R_i}\{1-\pi(X_i, Z_i, W_i;\alpha)\}^{1-R_i},$$
and $\hat{\beta}$ and $\hat{\gamma}$ are the weighted least-squares estimators
\begin{align*}
\sum_{i=1}^{m} B(X_i, Z_i, W_i)\{Y_i - g_\theta(X_i, Z_i, W_i;\beta)\}^2&=0,\\
\sum_{i=1}^{m} W_i(W_i-1)\{Y_i - C_\theta(X_i;\gamma)\}^2&=0,
\end{align*}
where $B(x,z,w)$ is any function of $x, z, w$ that has the same dimensionality as $\beta$. 

Hereafter, we put additional Conditions S1--S8 to obtain the desired theoretical results. However, we relegated these conditions to the supplementary material because they are the standard regularity conditions for estimation using the method of moments and empirical likelihood.

\subsection{Optimal estimator using the method of moments in Settings 1 and 2}
\label{sec:4.1}

Considering the orthogonality of the estimating equation relative to nuisance parameters, adaptive estimators using the method of moments are obtained by simply replacing the unknown functions with the estimated functions.

In Setting 1, we define a method of moments estimator $\hat{\theta}_{\MM 1}$ as the solution to
    \begin{align}
   S^{[1]}_{\eff, \theta}(\hat{\alpha},\hat{\beta}, \hat{\gamma})=n^{-1}\sum_{i=1}^N \llp \delta_i W_i  \hat{D}_\theta(R_i, X_i, Y_i, Z_i, W_i)  + (1-\delta_i W_i) \hat{C}_\theta(X_i)\rrp =0, \label{MM1}
    \end{align}
  whereas in Setting 2, we define $\hat{\theta}_{\MM 2}$ by the solution to an estimating equation obtained by replacing $\hat{C}_\theta(x)$ with $\hat{C}_\theta$.
We omit the proof of Theorem \ref{thm.4.1} because it follows from the standard asymptotic theory for the method of moments.

\begin{theorem}
\label{thm.4.1}
In Settings 1 and 2, we assume the informative sampling of definition \ref{d1}, Conditions \ref{c1}, \ref{c2}, and {\rm S}1--{\rm S}3. If all working models are correct, then two adaptive estimators $\hat{\theta}_{\MM 1}$ and $\hat{\theta}_{\MM 2}$ achieve the efficiency bound $\{E(S_{\eff, j}^{\otimes 2})\}^{-1}$ for each Setting $j=1,2$. If either $\pi(X, Z, W)$ or $g_{\theta}(X, Z, W)$ is correctly specified, $\hat{\theta}_{\MM 1}$ and $\hat{\theta}_{\MM 2}$ still have consistency and asymptotic normality with variances $E(\varphi_{\eff,1}^{\otimes 2})$ and $E(\varphi_{\eff,2}^{\otimes 2})$ for each Setting $j=1,2$, where
$
\varphi_{\eff, j} = [ E \{ \pd S_{\eff,j}(\theta; \alpha^*, \beta^*, \gamma^*)/\pd \theta^\top \}] ^{-1} S_{\eff,j}(\theta; \alpha^*, \beta^*, \gamma^*).
$
\end{theorem}

\subsection{Optimal estimator using empirical likelihood in Settings 1 and 2}
We consider empirical likelihood based estimators with the same asymptotic variances detailed in Theorem \ref{thm.4.1} for Settings 1 and 2. There are two primary motivations for considering the empirical likelihood methods. First, using the empirical likelihood method facilitates the easy derivation of multiple-robust estimators. Second, an elementary extension of the optimal empirical likelihood estimator from Setting 2 can provide the optimal estimator for Setting 3. Although it may be feasible to construct an estimator by the method of moments and extend it to Setting 3, we lean toward empirical likelihood estimators for their theoretically appealing properties; see Remark \ref{rem:2} for additional details.

One natural empirical likelihood estimator based on the efficient score \eqref{eff_M} is defined as$$\hat{\theta}_{\EL,\mathrm{Q}}=\argmax_\theta\argmax_{p_1,\dots, p_N} \sum_{i=1}^N \log p_i,$$
subject to $\sum_{i=1}^N p_i=1$, and
\begin{align*}
\sum_{i=1}^N p_i\delta _i W_i D_\theta(R_i, X_i, Y_i, Z_i, W_i)&=0,\\
\sum_{i=1}^N p_i(1-\delta_iW_i) C_{\theta}(X_i)&=0.
\end{align*}
By directly applying the theory in \citet*{qin09}, we can easily demonstrate that this maximum empirical likelihood estimator has desirable asymptotic properties. However, we do not adopt $\hat{\theta}_{\EL,\mathrm{Q}}$ as our estimator because the function $D_\theta(r,x,y,z,w)$ includes the two working models $\pi(x,z,w)$ and $g(x,z,w)$, making it intractable to construct a multiple-robust estimator. Let our candidate models for $\pi(x,z,w)$ and $g_\theta(x,z,w)$ be  $\pi^{[j]}(x,z,w)$ $(j=1,\dots. J)$ and $g_\theta^{[k]}(x,z,w)$ ($k=1,\dots,K$), respectively. Then, $\hat{\theta}_{\EL,\mathrm{Q}}$ requires $JK$ constraints to obtain multiple robustness. Hereafter, in Settings 1 to 3, we propose two-step maximum empirical likelihood estimators to realize multiple robustness under only $J+K$ constraints. We also present the use of $C^{[l]}_\theta$ $(l=1,\dots, L)$ working models for $C_\theta(x)$ to achieve the efficiency bound.

The first-step empirical likelihood weights are common in all settings. We define the maximum empirical weights in the first step as
$$\hat{\theta}_1=\argmax_{\theta}\argmax_{p^{(1)}_1,\dots, p^{(1)}_{m}} \sum_{i=1}^{m} \log p^{(1)}_i,$$
subject to $\sum_{i=1}^{m} p^{(1)}_i=1$,  and 
\begin{align*}
\sum_{i=1}^{m} p^{(1)}_i \{\hat{\pi}^{[j]}(X_i, Z_i, W_i)-\bar{\pi}_n^{[j]}\}&=0, \quad (j=1,\dots, J)\\
\sum_{i=1}^{m} p^{(1)}_i \{ W_i \hat{g}^{[k]}_\theta(X_i, Z_i, W_i)-\bar{g}^{w[k]}_{\theta}\}&=0, \quad (k=1,\dots, K)
\end{align*}
where $\bar{\pi}^{[j]}_n$ and $\bar{g}^{w[k]}_{\theta}$  are defined as
$\bar{\pi}^{[j]}_n = n^{-1}\sum_{i=1}^n \hat{\pi}^{[j]}(X_i, Z_i, W_i)$ and $\bar{g}^{w[k]}_\theta = n^{-1}\sum_{i=1}^{n} W_i \hat{g}^{[k]}_\theta(X_i, Z_i, W_i)$, respectively.
The first-step maximum empirical weights are denoted as
$(\hat{p}^{(1)}_1,\dots, \hat{p}^{(1)}_{m})=(\hat{p}^{(1)}_1(\hat{\theta}_1),\dots, \hat{p}^{(1)}_{m}(\hat{\theta}_1)).$

Next, we define the maximum empirical likelihood weights in the second step as
$$\hat{\theta}_{2}=\argmax_{\theta} \argmax_{p^{(2)}_1,\dots, p^{(2)}_N}\sum_{i=1}^N  \log p^{(2)}_i,$$
subject to $\sum_{i=1}^N p^{(2)}_i=1$, and 
\begin{align*}
\sum_{i=1}^N p^{(2)}_i \lp 1-\delta_iW_i\rp \hat{C}^{[l]}_\theta(X_i)=0. \quad (l=1,\dots, L)
\end{align*}
The second-step maximum empirical weights are denoted as
$(\hat{p}^{(2)}_1,\dots, \hat{p}^{(2)}_{m})=(\hat{p}^{(2)}_1(\hat{\theta}_2),\dots, \hat{p}^{(2)}_{m}(\hat{\theta}_2)).$
Then, we define the final maximum empirical likelihood estimator $\hat{\theta}_{\EL 1}$ in Setting 1 as the unique solution to
\begin{align}
\sum_{i=1}^N \hat{p}^{(2)}_i\hat{p}^{(1)}_i\delta_i R_i  W_i  U_\theta(X_i, Y_i)=0. \label{EL1}
\end{align}

\begin{remark}
\label{rem:1}
\noindent
  The first step of the empirical weights described above is a generalization of the multiple-robust estimator in \citet{han14}. Indeed, $ W_i= 1\,(i=1,\dots, n)$ yields the same empirical weights, implying that multiplication by $W_i$ is required to adjust the sampling bias under informative sampling.
  The concept for the second step of the empirical weights comes from \citet*{qin09}'s estimator $\hat{\theta}_{\EL, \mathrm{Q}}$. Therefore, our empirical likelihood weights are obtained by combining the ideas of \citet{han14} and \citet*{qin09}.
\end{remark}

\begin{remark}
\label{rem:2}
\noindent
 Our empirical likelihood estimator does not directly use $1/\hat{\pi}(x,z,w)$, unlike the method of moments estimator. Therefore, the finite-sample performance of the empirical likelihood estimator is better than that of the method of moments estimator when some $\hat{\pi}(x,z,w)$ can take on values near zero. See \citet{han14} for additional details.
\end{remark}

     In Setting 2, we must modify the empirical weights in the second step and define the maximum empirical likelihood estimator as 
     $$\hat{V}=\argmax_{V} \argmax_{p^{(2)}_1,\dots, p^{(2)}_n}\sum_{i=1}^n  \log p^{(2)}_i + (N-n)\log (1-V),$$ subject to $\sum_{i=1}^n p^{(2)}_i=1$, and $\sum_{i=1}^n p^{(2)}_i ( 1/W_i- V)=0$. Then, we define the empirical weights as $(\hat{p}^{(2)}_1,\dots, \hat{p}^{(2)}_{n})=(\hat{p}^{(2)}_1(\hat{V}),\dots, \hat{p}^{(2)}_{n}(\hat{V}))$.
Our maximum empirical likelihood estimator $\hat{\theta}_{\EL 2}$ in Setting 2 is defined as the unique solution to 
\begin{align}
\sum_{i=1}^N \hat{p}^{(2)}_i\hat{p}^{(1)}_i\delta_i R_i   U_\theta(X_i, Y_i)=0. \label{EL2}
\end{align}
In Setting 2, modeling $C_\theta$ and multiplying by the sampling weights in \eqref{EL2} are unnecessary because such information is already carried by the second term in the empirical likelihood.
Then, our estimators $\hat{\theta}_{\EL 1}$ and $\hat{\theta}_{\EL 2}$ have desired multiple robustness.

\begin{theorem}
\label{thm.4.2}
In Settings 1 and 2, we assume the informative sampling of definition \ref{d1}, Conditions \ref{c1}, \ref{c2}, and {\rm S}4--{\rm S}7. If each of the $J$ models for the response mechanism, $K$ models for the regression function, and $L$ models for $C_\theta(x)$ include the correct model, then the two adaptive estimators $\hat{\theta}_{\EL 1}$ and $\hat{\theta}_{\EL 2}$ achieve the efficiency bound $\{E(S_{\eff, j}^{\otimes 2})\}^{-1}$ for each Setting $j=1,2$. If at least one of the $J+K$ models for the response and outcome regression models is correctly specified, then $\hat{\theta}_{\EL 1}$ and $\hat{\theta}_{\EL 2}$ still have consistency.
\end{theorem}


\subsection{Optimal estimator using empirical likelihood in Setting 3}
Finally, we propose the most efficient estimator in Setting 3 by extending the estimator in Setting 2. Recall that $\tau=\tau(P_1)$ is the target parameter in the external data, where $P_1$ is the distribution of external data. Suppose that an estimator $\tilde{\tau}$ for $\tau$, its variance estimator $\tilde{\Sigma}_1$, and the sample size of the external data $N_1$ are available from the external source. Additionally, suppose that we can access summary statistics such as (i) $Z$-estimator (the solution to $E\{U_\tau(X,Z)\}=0$), (ii) regression coefficients $E(X\mid Z;\tau)$, and (iii) conditional density $f(x\mid z;\tau)$. According to Condition \ref{c3}, because there is no missingness for $X$ and $Z$ after sampling, the optimal estimator is obtained by solving $ 
n^{-1}\sum_{i=1}^N \llp \delta_i W_i  D^*_\tau(X_i, Z_i)  + (1-\delta_i W_i) C^*_\tau(X_i)\rrp =0,
$
where $D^*_\tau$ and $C^*_\tau$ are defined in Theorem 3.1 in  \citet*{morikawa22} dependent on the target parameter $\tau$. For example, if a $Z$-estimator is of interest, then $D^*_\theta = U_\tau(X,Z)$ and $C^*_\tau=E\{(W-1)U_\tau(X,Z)\mid X\}/E(W-1\mid X)$.

Then, by using the optimal score function for $\tau$, the maximum empirical likelihood estimator in Setting 3 in the second step is defined through the maximizer of $(\tau, V)$ in
$$\argmax_{p^{(2)}_1,\dots, p^{(2)}_n}\sum_{i=1}^n  \log p^{(2)}_i + (N-n)\log (1-V)\ - 2^{-1} N_1 ( \tilde{\tau} - \tau )^{\top} \tilde{\Sigma}_1^{-1} (\tilde{\tau} - \tau)^{\top},$$ 
subject to $\sum_{i=1}^n p^{(2)}_i=1$, $\sum_{i=1}^n p^{(2)}_i ( 1/W_i- V)=0$, and $\sum_{i=1}^n p_i^{(2)} D^*_\tau(X_i, Z_i)=0$.
We incorporate the information on $\tilde{\tau}$ into the likelihood to leverage our prior knowledge that $\tilde{\tau}$ is asymptotically normally distributed with mean $\tau$ and variance $\tilde{\Sigma}_1$, as discussed in \citet{zhang2020generalized}. Furthermore, the third constraint above is essential for efficiently estimating $\tau$ using the internal data. Then, our final estimator $\hat{\theta}_{\EL 3}$ is obtained by solving \eqref{EL2} with respect to $\theta$.
Note that the external summary statistics contribute to the empirical likelihood through the term $N_1 (\tilde{\tau} - \tau)^{\top} \tilde{\Sigma}_1^{-1} (\tilde{\tau} - \tau)$. When $N_1$ is small or $\tilde{\Sigma}_1$ is large, this term exerts minimal influence on the empirical objective, limiting the efficiency gains. Therefore, sufficient precision of the external information is essential for improving performance under no heterogeneity between the internal and external data.
Under appropriate regularity conditions, the proposed estimator is multiply robust and attains the efficiency bound associated with incorporating external summary statistics.



\begin{theorem}
\label{thm:4.3}
In Setting 3, we assume the informative sampling of definition \ref{d1}, Conditions \ref{c1}--\ref{c6}, {\rm S}4--{\rm S}6, and {\rm S}8. If each of the $J$ models for the response mechanism and $K$ models for the regression function include the correct model, the adaptive estimator $\hat{\theta}_{\EL 3}$ achieves the efficiency bound addressed in \S 3.2. If at least one of the $J+K$ models for the response and outcome regression models is correctly specified,  $\hat{\theta}_{\EL 3}$ still has consistency.
\end{theorem}




\section{Numerical Study}
\label{sec:5}
We conducted numerical studies to investigate the performance of our proposed estimators for $\theta = E(Y)$ or $U_\theta(y)=y-\theta$ in Settings 1 to 3. The datasets were identically and independently generated according to the following distribution: $X\sim N(0,\,1/2)$, $Z\sim N(0,\,1/2)$, $Y\mid (x, z)\sim N(x-z,\,1/2)$, $\log (W-1)\mid (x, y, z) \sim N(2.95-0.25x-0.45y-0.1z, 0.05)$,  $P(\delta = 1 \mid x,y,z,w)=w^{-1}$, and $P(R=1\mid \delta = 1, x,z,w) = \mathrm{expit}(-0.3+0.75 x - 0.5z+0.05 w)$, where $\mathrm{expit}(\cdot)$ is the inverse function of the logistic function. The above data generation process with a population size $N=10,000$ yielded the sample size of $n\approx 600$ and the observed sample size of $m\approx 400$. We iteratively estimated $\theta = E(Y)$ 2,000 times.
Under this model setup, the true function of $E(Y\mid \delta=1, x, z, w)$ is a linear function of $x$ and $z$, $\log(w-1)$ becomes a linear function of $x$, and $C_\theta(x)$ becomes a linear function of $x$ and $\theta$.
We assume that $C_\theta(x)$ is correctly specified and investigate the double or multiple robustness of our estimators for the specification of $\pi(x,z,w)$ or $g_\theta(x,z,w)$.
In practice, if there are concerns about potential model misspecification of $C_\theta(x)$, we recommend using the approach of specifying multiple candidate models as demonstrated in Section 4.

\subsection{Setting 1.}
We compared the Horvitz-Thompson (HT) and Kim and Haziza (KH) estimators given in \eqref{HT} and \eqref{KH} to our proposed multiple-robust empirical likelihood estimators. In this numerical study, $C_\theta(x)$ was correctly specified, and we prepared three models for each $\pi(x,z,w)$ and $g_\theta(x,z,w)$, resulting in a total of six working models: (1) (Correct) $\pi^{(1)}(x,z,w;\alpha^{(1)}) = \mathrm{expit}(\alpha^{(1)}_0+\alpha^{(1)}_1 x + \alpha^{(1)}_2 z + \alpha^{(1)}_3 w)$; (2) $\pi^{(2)}(x,z,w;\alpha^{(2)}) =  \mathrm{expit}(\alpha^{(2)}_0+\alpha^{(2)}_1 x + \alpha^{(2)}_2 z + \alpha^{(2)}_3 xz)$,
    (3) $\pi^{(3)}(x,z,w;\alpha^{(3)}) =  \mathrm{expit}\{s_1(x;\alpha^{(3)}) + s_2(z;\alpha^{(3)}) + s_3(\log(w-1);\alpha^{(3)}) \}$, (4) (Correct) $g^{(1)}_\theta(x,z,w;\beta^{(1)}) = \beta^{(1)}_0 + \beta^{(1)}_1 x + \beta^{(1)}_2 z + \beta^{(1)}_3 \log(w-1)-\theta$,  (5) $g^{(2)}_\theta(x,z,w;\beta^{(2)}) = \beta^{(2)}_0 + \beta^{(2)}_1 x + \beta^{(2)}_2 z + \beta^{(2)}_3 xz-\theta$, (6) $g^{(3)}_\theta(x,z,w;\beta^{(3)}) = \tilde{s}_1(x;\beta^{(3)}) + \tilde{s}_2(z;\beta^{(3)}) +\tilde{s}_3(w;\beta^{(3)})-\theta$,
    where $s_j(\cdot)$ and $\tilde{s}_j(\cdot)\,(j=1,2,3)$ are smoothing splines with three degrees of freedom. It should be noted that $\pi^{(1)}$ and $g^{(1)}$ are correctly specified models, whereas $\pi^{(2)}$ and $g^{(2)}$ are misspecified because the sampling weights are not included in the models. The generalized additive models $\pi^{(3)}$ and $g^{(3)}$ are also misspecified, but it is expected that these models will achieve better performance because they contain information on sampling weights. 
    
   We compared $\mathrm{HT}_i$ (Horvitz-Thompson), $\mathrm{KH}_{ij}$ (Kim and Haziza), and our proposed double-robust estimators to the methods of moments estimator $\mathrm{MM}_{ij}$ and our proposed multiple-robust empirical likelihood estimators $\mathrm{EL}_{ij|kl}$, where $i,j,k,l \in \{0,1\}$. The estimators $\mathrm{HT}_1$ and $\mathrm{HT}_0$ are the Horvitz-Thompson estimators with $\pi^{(1)}$ and $\pi^{(2)}$. The estimators $\mathrm{KH}_{11}$, $\mathrm{KH}_{10}$, $\mathrm{KH}_{01}$, and $\mathrm{KH}_{00}$ are the Kim and Haziza estimators with $(\pi^{(1)}, g^{(1)}_\theta)$, $(\pi^{(1)}, g^{(2)}_\theta)$, $(\pi^{(2)}, g^{(1)}_\theta)$, and $(\pi^{(2)}, g^{(2)}_\theta)$. We prepared four models for our empirical likelihood based multiple-robust estimators: $\mathrm{EL}_{10|10}$ (combination of $\pi^{(1)}, \pi^{(2)}, g^{(1)}_\theta, g^{(2)}_\theta$), $\mathrm{EL}_{10|00}$ (combination of $\pi^{(1)}, \pi^{(2)}, g^{(2)}_\theta, g^{(3)}_\theta$), $\mathrm{EL}_{00|10}$ (combination of $\pi^{(2)}, \pi^{(3)}, g^{(1)}_\theta, g^{(2)}_\theta$), and $\mathrm{EL}_{00|00}$ (combination of $\pi^{(2)}, \pi^{(3)}, g^{(2)}_\theta, g^{(3)}_\theta$).
    
    The left panel in Figure \ref{fig:2} presents a boxplot of the results of the estimators for $E(Y)$. The misspecified HT estimator is heavily biased when the response model is wrong, and it has consistency but is less efficient even when the response model is correct. Both KH and proposed estimators have double robustness, but the proposed estimators are less biased and more efficient when both models do not contain the correct models because they include nearly-correct misspecified models $\pi^{(3)}$ and $g_\theta^{(3)}$. 
   
\subsection{Settings 2 and 3.}
\begin{figure}
    \centering
		  \includegraphics[width=140mm]{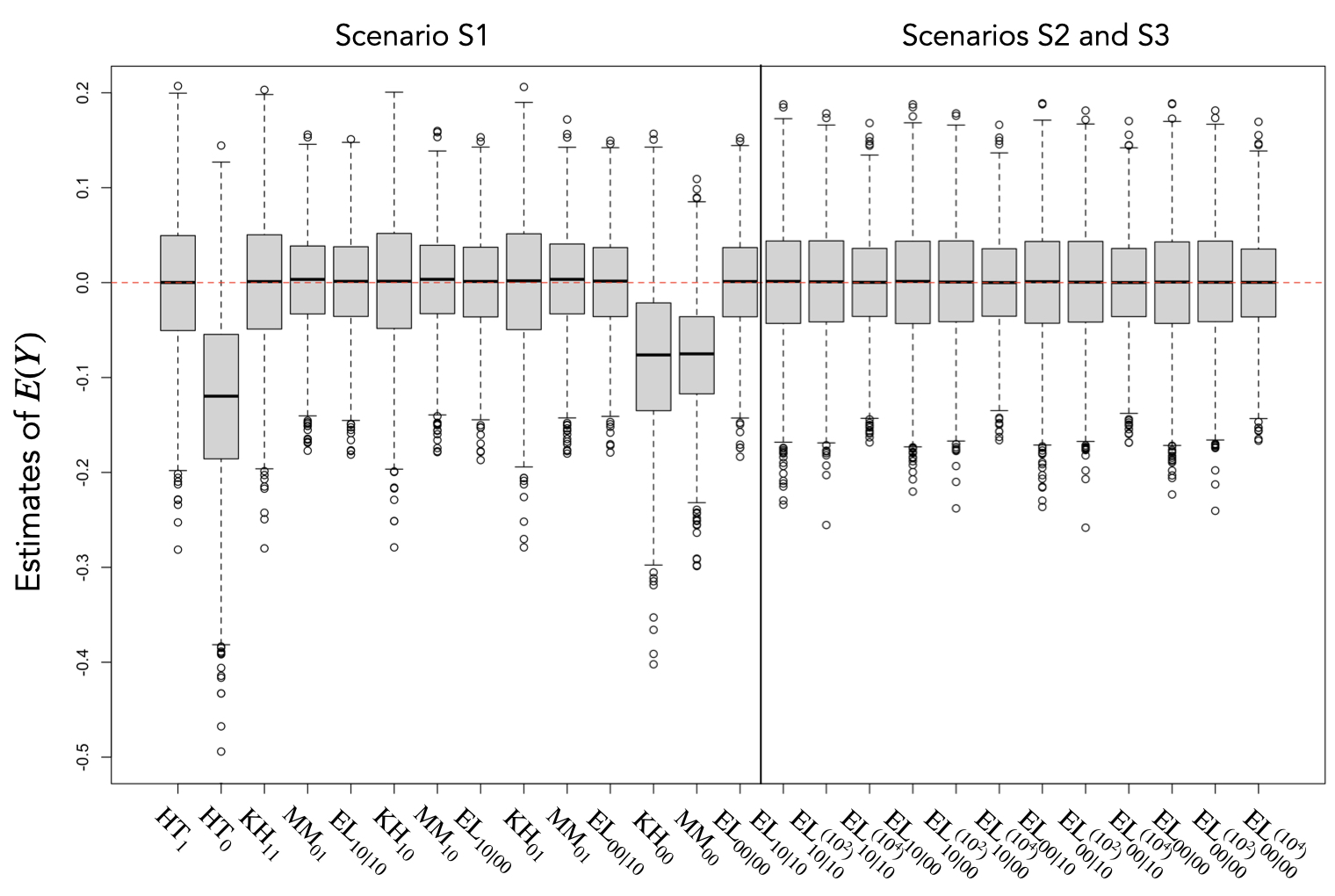}
		 \caption{(Left) Results in Setting 1: $\mathrm{HT}_i$ (Horvitz-Thompson), $\mathrm{KH}_{ij}$ (Kim and Haziza), proposed methods of moments double-robust estimators $\mathrm{MM}_{ij}$ (methods of moments), and proposed empirical likelihood based multiple-robust estimators $\mathrm{EL}_{ij|kl}$, where $i,j,k,l \in \{0,1\}$. The indices $i$ and $j$, and $k$ and $l$ take on values of one if the two working models are correct and values of zero otherwise. (Right) Results in Settings 2 and 3: $\mathrm{EL}_{ij|kl}\,(i,j,k,l \in \{0,1\})$ in Setting 2 and $\mathrm{EL}_{ij|kl}^{(N_1)}\,(i,j,k,l \in \{0,1\})$ in Setting 3 with the sample size of the external source being $N_1=100$ or $10,000$.}    \label{fig:2}
\end{figure}

    We assumed that $N_1^{-1}\sum_{i=1}^{N_1} X_i$ is available from an external data source with the sample size of the data source being $N_1=100$ or $10,000$.  
   The right panel in Figure \ref{fig:2} presents a boxplot of the results of the proposed empirical likelihood based estimators with $\mathrm{EL}_{ij|kl}\,(i,j,k,l \in \{0,1\})$ in Setting 2 and $\mathrm{EL}_{ij|kl}^{(N_1)}\,(i,j,k,l \in \{0,1\})$ in Setting 3 with the sample size of the external source being $N_1$=100 or 10,000. Our proposed estimators in Settings 2 and 3 still have multiple robustness. Furthermore, as the external information increases, the efficiency of the proposed estimators increases. A comparison of the two estimators $\mathrm{EL}_{10|10}$ and $\mathrm{EL}^{(10^4)}_{10|10}$ indicates that rich information on covariates in the external data can facilitate the creation of more efficient estimators compared to sparse information on covariates in the internal data.

We used 200 bootstrap samples to estimate the variance of the estimators. The coverage rates for the 95\% confidence intervals for $\mathrm{EL}_{10|10}$ in Settings 1, 2, and 3 when $N_1=100$ and $N_1=10,000$ were 95.05\%, 96.65\%, 96.25\%, and 96.55\%, respectively. Therefore, variance estimation with bootstrap samples is acceptable.

\section{Real Data Analysis}

We applied our methods to data from the 2015-2016 National Health and Nutrition Examination Survey (NHANES), which is a program of studies designed to assess the health and nutritional status of adults and children in the United States. The dataset analyzed in this section is accessible at \url{https://wwwn.cdc.gov/nchs/nhanes/continuousnhanes/default.aspx?BeginYear=2015}. Following \citet*{chen21}, \citet*{duan10}, and \citet{parker23}, we analyzed the NHANES data without assuming the non-informativeness of the sampling mechanism. 

The body mass index (BMI) is a simple and useful measure for investigating the degree of obesity in individuals. However, the BMI is not a direct measure of body fat. Instead of using the BMI, dual-energy X-ray absorptiometry (DXA) has been accepted as an alternative method for measuring body fat. In this study, we focused on estimating the sample mean of DXA.
Let $Y$ be the value of DXA ($\%$), $X$ be BMI 
($\mathrm{g}/\mathrm{m}^2$), and $Z$ be age (years). It is implausible that the absence of DXA scans is driven by the DXA values themselves, which are not routinely available to the general public; rather, it is more natural to attribute the lack of DXA uptake to other covariates such as BMI and age. Accordingly, following \citet*{duan10}, we assume PMAR.

Since only a small number of units of $X$ were missing, we removed those data from our analysis, adjusted the sampling weights, and created a dataset that is consistent with Setting 2. The counts of the finite population, sampled units, and observed units are $N=316,481,044$, $n=8,756$, and $m=4,590$. We also utilize additional information on $X$ from the National Health Interview Survey (NHIS). The NHIS site (\url{https://www.cdc.gov/nchs/nhis/shs/tables.htm}) reports that the point estimate for the obesity rate of $\{X\geq 30\}$ in the NCHS in 2015 was $\tilde{\tau}=0.295$ with an estimated standard deviation $\tilde{\sigma}_1=0.2057$ and sample size $N_1=33,672$. 

We estimated the population mean of dual-energy X-ray absorptiometry using four different estimators: $\mathrm{HT}_1$, $\mathrm{HT}_3$, $\mathrm{EL}_{13|13}$, and $\mathrm{EL}_{13|13}^{(N_1)}$, where $\mathrm{HT}_1$ and $\mathrm{HT}_3$ are the Horvitz-Thompson estimators employing response models $\pi^{(1)}$ and $\pi^{(3)}$, as detailed in \S \ref{sec:5}. $\mathrm{EL}_{13|13}$ represents our proposed empirical likelihood estimator for Setting 2, which incorporates $\pi^{(1)}$, $\pi^{(3)}$, and $g_\theta^{(1)}$ from \S \ref{sec:5},  and $\tilde{g}^{(3)}_\theta(x,z,w;\beta^{(3)}) = \tilde{s}_1(x;\beta^{(3)}) + \tilde{s}_2(z;\beta^{(3)}) +\tilde{s}_3(\log(w-1);\beta^{(3)})-\theta$. Finally, $\mathrm{EL}_{13|13}^{(N_1)}$ is our proposed estimator for Setting 3 utilizing the same working models.

The point estimates (along with their standard errors) by $\mathrm{HT}_1$, $\mathrm{HT}_3$, $\mathrm{EL}_{13|13}$, and $\mathrm{EL}_{13|13}^{(N_1)}$ were calculated as 31.685 (0.168), 31.601 (0.215), 31.845 (0.206), and 31.654 (0.197). These values suggest that the true response mechanism $\pi(x,z,w)$ may be essentially missing completely at random because all the estimates are closely aligned. Furthermore, the estimator $\mathrm{EL}_{13|13}^{(N_1)}$ shifted slightly downward to the estimate in NHIS ($\tilde{\tau}=0.295$) compared to $\mathrm{EL}_{13|13}$, and its standard error decreased, likely as a result of the incorporation of additional information from an external data source.

\section{Discussion}

This study concentrated on the effective application of sampling weights for the analysis of sampled data with nonresponse. While our estimators are widely applicable, they are not without limitations. Firstly, Conditions \ref{c1} and \ref{c2}, addressing the ignorability of the response mechanism, may be restrictive for practical research. As a result, it is essential to refine our methods to accommodate non-ignorable or missing not at random response mechanisms. Secondly, our methodology requires the external information to be completely sampled, which means that any parameters dependent on the outcome variable ($y$) are excluded from our current framework. Exploring ways to modify our techniques to include such parameters is a valuable avenue for future research. Thirdly, this study presupposes the background superpopulation model. Exploring a finite population asymptotic framework is a pressing necessity. Furthermore, advancing our optimal estimators to embrace Bayesian empirical likelihood represents a significant practical endeavor \citep{rao2010, zhao2020}.
Finally, our current framework assumes Poisson sampling, which allows us to use standard independent and identically distributed assumptions in the semiparametric theory.
Extending our methodology to accommodate more general sampling designs, such as stratified or unequal probability sampling, introduces significant theoretical challenges due to the inherent non-i.i.d. nature of the data.
Foundational work by \cite{mcneney2000application} has developed semiparametric theory under non-i.i.d. settings and established the convolution theorem that provides a semiparametric efficiency bound.
Building upon such frameworks to incorporate complex sampling designs into our setting remains an important and challenging direction for future research.



\appendix

\section*{Supplementary material}
\label{SM}

\section{Additional regularity conditions}

\begin{enumerate}[label=(S\arabic*), ref=(S\arabic*)]
    \item The estimator $(\hat{\alpha}^{\top}, \hat{\beta}^{\top}, \hat{\gamma}^{\top})^{\top}$ converge to $({\alpha^*}^{\top}, {\beta^*}^{\top}, {\gamma^*}^{\top})$ in probability and have asymptotic normality, where the asymptotic variance is nonsingular.

     \item For each $j=1,2$, $\partial S^{[j]}_{\eff, \theta}(\alpha, \beta, \gamma)/\partial\lp \theta^{\top}, {\alpha}^{\top}, {\beta}^{\top}, {\gamma}^{\top} \rp$ is continuous at $({\theta^*}^{\top}, {\alpha^*}^{\top}, {\beta^*}^{\top}, {\gamma^*}^{\top})$ with probability one, and there is a neighborhood $\mathcal{N}$ of $({\theta^*}^{\top}, {\alpha^*}^{\top}, {\beta^*}^{\top}, {\gamma^*}^{\top})$ such that
   \begin{align*}
       E\llp \sup_{ ({\theta}^{\top}, {\alpha}^{\top}, {\beta}^{\top}, {\gamma}^{\top}) \in \mathcal{N} }
       \lnorm
       \frac{\partial S_{\eff, \theta}( \alpha, \beta, \gamma)}{\partial\lp \theta^{\top}, {\alpha}^{\top}, {\beta}^{\top}, {\gamma}^{\top} 
       \rp } 
       \lnorm
       \rrp 
       <\infty.
   \end{align*}

   \item For each $j=1,2$, $S^{[j]}_{\eff,\theta}(\alpha, \beta, \gamma)$ is continuously differentiable at all $(\theta, \alpha, \beta, \gamma)$ with probability one and is bounded by some integrable function.

   \item Let $\alpha_{j}\,(j=1,\dots, J)$, $\beta_{k}\,(k=1,\dots, K)$ and $\gamma_{l}\,(l=1,\dots, L)$ be the parameter of candidate working models $\pi^{[j]}$, $g^{[k]}$, and $C^{[l]}(X)$.
    We use $\hat{\alpha}_{k}$, $\hat{\beta}_{j}$, and $\hat{\gamma}_{l}$ to denote the estimators of $\alpha_{k}$, $\beta_{j}$, and $\gamma_{l}$.
    $N^{1/2}(\hat{\alpha}_k - \alpha_k^*)$, $N^{1/2}(\hat{\beta}_j - \beta_j^*)$, and $N^{1/2}(\hat{\gamma}_l - \gamma_l^*)$ are bounded in probability, where $\alpha_k^*$, $\beta_j^*$, and $\gamma_l^*$ are the probability limit of $\hat{\alpha}_k$, $\hat{\beta}_j$, and $\hat{\gamma}_l$. Let  $\alpha^*=({\alpha_1^*}^{\top}, \ldots, {\alpha_J^*}^{\top})^{\top}$, $\beta^*=({\beta_1^*}^{\top}, \ldots, {\beta_K^*}^{\top})^{\top}$, and $\gamma^*=({\gamma_1^*}^{\top}, \ldots, {\gamma_L^*}^{\top})^{\top}$.

    \item $E(W^3)<\infty$.

    \item Suppose that $\pi^{[1]}(x,z,w;\alpha_1)$ and $g^{[1]}(x,z,w;\beta_1)$ are correctly specified. Then, functions $\|R \delta W(1-\delta W) U_{\theta}/\pi^{[1]} \|$, \\ $\|\partial^2 \{\hat{h}^{\top}({\alpha}, {\beta})/{\pi}^{[1]}\}/ \partial (\alpha^{\top}, \beta^{\top}) \partial (\alpha^{\top}, \beta^{\top})^{\top}\|$, $\|\hat{h}^{\top}({\alpha}, {\beta})/{\pi}^{[1]}\|^3$, \\$\|\partial \{\hat{h}^{\top}({\alpha}, {\beta})/{\pi}^{[1]}(X, Z, W;{\alpha}_1)\}/ \partial(\alpha^{\top}, \beta^{\top})\|$, $\|R\delta W U_{\theta}(X,Y)/\pi^{[1]}\|$, $\|(1 + \hat{\rho}^{\top}\hat{h})^{-1} R\delta W \llp  U_{\theta^*}(X, Y) - g^{[1]}\rrp \|$, $\|(1 - \delta W)^2 \mathcal{C}_{\theta}(X; \gamma)^{\otimes 2} \|$, $ \| R\hat{h}^{\otimes 2}/\{ (\pi^{[1]})^2 W\}  \|$, $ \|\delta R  (\pi^{[1]})^{-1}  \partial \hat{h}/\partial \alpha_1^{\top}  \|$, $ \|\delta R  (\pi^{[1]})^{-2}  \hat{h} \partial \pi^{[1]}/ \partial \alpha_1  \|$, $ \|\delta R \hat{h}/ \pi^{[1]} \|$, $\|R\delta W U_{\theta}(X,Y) \hat{h}^{\top}/(\pi^{[1]})^2 \|$, $\|R\delta W (W-1)U_{\theta}(X, Y)  \mathcal{C}_{\theta}(X;\gamma)^{\top}/\pi^{[1]}\|$, $\|R\delta W (\pi^{[1]})^{-2} U_{\theta}(X, Y) {\partial \pi^{[1]}}/{\partial \alpha_1^{\top}} \|$, and $\| R \delta W/\pi^{[1]}
\{\partial U_{\theta}(X,Y)/\partial \theta^{\top}
\}
\|$ are continuous and bounded by some integrable function in the neighborhood of $({\alpha^*}^{\top}, {\beta^*}^{\top}, {\gamma^*}^{\top}, {\theta^*}^{\top})$, where $\hat{h}({\alpha}, {\beta},{\theta})^{\top}=( {\pi}^{[1]}({\alpha}_1)-\bar{\pi}_n^{[1]}({\alpha}_1), \ldots, {\pi}^{[J]}({\alpha}_J)-\bar{\pi}_n^{[J]}({\alpha}_J), W {g}_{{\theta}}^{[1]}( {\beta}_1)^{\top}-\bar{g}^{w[1]}_{{\theta}}({\beta}_1)^{\top},\\ \ldots,  W {g}^{[K]}({\beta}_K)^{\top}-\bar{g}^{w[K]}_{{\theta}} ({\beta}_K)^{\top} )$, $\rho^*$ is the probability limit of the Lagrange multipliers $\hat{\rho}$ satisfied \eqref{nonchange_ELweight}, and $\mathcal{C}_{{\theta} }(X; {\gamma})=( {C}^{[1]}_{ {\theta} }(X; {\gamma}^{[1]})^{\top}, \ldots, {C}^{[L]}_{ {\theta} }(X; {\gamma}^{[L]})^{\top} )^{\top}$. 

\item Functions
        $\|( 1-\delta W ) \mathcal{C}_{{\theta} }(X; {\gamma})\|^3$, $\|\partial \{( 1-\delta W ) \mathcal{C}_{{\theta} }(X; {\gamma})\}/ \partial(\theta^{\top}, \gamma^{\top})\|$, and $\|\partial^2 \{( 1-\delta W ) \mathcal{C}_{{\theta} }(X; {\gamma})\}/ \partial (\theta^{\top}, \gamma^{\top}) \partial (\theta^{\top}, \gamma^{\top})^{\top}\|$
     are continuous and bounded by some integrable function with respect to the probability distribution $(X, Z, W)$ in the neighborhood of $({\gamma^*}^{\top},  {\theta^*}^{\top})$.

     \item Functions
        $\|\partial^2 \{WD_{\tau}^*(X_i, Z_i)\}/ \partial \tau^{\top} \partial \tau\|$, $\|\partial \{WD_{\tau}^*(X_i, Z_i)\}/ \partial\tau^{\top}\|$, and $\|WD_{\tau}^*(X_i, Z_i)\|^3$ are continuous and bounded by some integrable function in the neighborhood of $\tau^*$.
     Functions\\ $\|\delta R W^2 U_{\theta}(X, Y) D^*_\tau(X, Z)^{\top}/\pi^{[1]} \|$, $\| R \delta W/\pi^{[1]}  ¥\{\partial U_{\theta}(X,Y)/\partial \theta^{\top}
     \}
     \|$, $\| (W-1)D_{\tau}^*\|$, $\| W(D_{\tau}^*)^{\otimes2}\|$, and $\| \partial D_{\tau}^*/\partial \tau\|$ are bounded by some integrable function in the neighborhood of $({\alpha_1^*}^{\top}, {\theta^*}^{\top}, {\tau^*}^{\top})$.   
\end{enumerate}

\section{Technincal proofs}
Proof of Theorem 1.
Let $\Lambda_2\subset \mathcal{H}$ be a functional space such that 
$$\Lambda_2 = \llp h\in\mathcal{H}\;\bigg|\; E(h\mid x,y,z,w)=0 \rrp.$$
It follows from Theorem 9.2 in \citet{tsiatis2006} that
\begin{align*}
\Lambda_2 &= \llp (1-\delta W)g_1(X) + \delta W \lp 1-\frac{R}{\pi} \rp g_2(X,Z,W) \;\bigg|\; \mathrm{for~any~}g_1, g_2 \rrp.
\end{align*}
Noting that $B_\theta U_\theta(X, Y)\in \Lambda^{F\perp}$, Theorem 10.1 in \citet{tsiatis2006} yields that the efficient influence function is obtained by computing the projection
$$\varphi_{\eff}=\prod\lp \frac{\delta WR}{\pi}B_\theta U_\theta(X,Y) \;\bigg|\; \Lambda_2^{\perp}\rp=B_\theta S_{\eff,\theta},$$
where $B_\theta$ is defined in Theorem 1 and 
$$
S_{\eff,\theta}=\prod\lp \frac{\delta WR}{\pi}U_\theta(X,Y) \;\bigg|\; \Lambda_2^{\perp}\rp.
$$
Therefore, it remains to find the efficient score $S_{\eff,\theta}$, i.e., $g_1^*(X), g_2^*(X,Z,W)$ satisfying, for any $g_1(X)$ and $g_2(X,Z,W)$,
\begin{align*}
& E\bigg[ \llp \frac{\delta W R}{\pi}U_\theta(X,Y) - (1-\delta W)g^*_1-\delta W \lp 1- \frac{R}{\pi}\rp g^*_2 \rrp\\
&\times \llp  (1-\delta W)g_1 + \delta W \lp 1 - \frac{R}{\pi}\rp g_2 \rrp  \bigg] = 0.
\end{align*}

By simple algebra, we have, for any $g_1$ and $g_2$,
\begin{align*}
E\lllp \llp (1-W)U_\theta(X,Y)  + (1-W)g_1^*  \rrp g_1 \rrrp - E\lllp \llp WO U_\theta(X,Y) + WOg_2^* \rrp g_2\rrrp
= 0,
\end{align*}
where $O=\pi^{-1}-1$, implying that
$$
g_1^* = - \frac{ E\llp (W-1)U_\theta(X,Y) \mid X \rrp}{E(W-1\mid X)},\quad
g_2^* = - g_\theta(X,Z,W).
$$
As a result, the efficient score can be written as
\begin{align*}
S_\eff &= \frac{\delta W R}{\pi} U_\theta(X,Y)  +  (1-\delta W)\frac{E\llp (W-1)U_\theta(X,Y) \mid X \rrp}{E(W-1\mid X)} + \delta W \lp 1-\frac{R}{\pi} \rp g_\theta(X,Z,W)\\
&= \delta W D_\theta(R,X,Y,Z,W) +  (1-\delta W)C_\theta(X),
\end{align*}
as desired.

\vspace{2ex}

\noindent 
Proof of Theorem 4.
For each Setting 1 and 2, we show multiple robustness and then prove the semiparametric efficiency. Suppose that $\pi^{[1]}$ is the correct model.

By using the method of Lagrange multipliers, the first-step empirical weights are obtainable as
\begin{align*}
    \hat{p}_i^{(1)}=m^{-1} \frac{1}{ 1 + \hat{\rho}^{\top} \hat{h}_i(\hat{\alpha}, \hat{\beta},\hat{\theta}_1) },\quad (i=1,\dots, m) 
\end{align*}
where 
\begin{align*}
    \hat{h}_i(\hat{\alpha}, \hat{\beta},\hat{\theta}_1)=&\left( \hat{\pi}^{[1]}(X_i, Z_i, W_i;\hat{\alpha}_1)-\bar{\pi}_n^{[1]}(\hat{\alpha}_1), \ldots, \hat{\pi}^{[J]}(X_i, Z_i, W_i;\hat{\alpha}_J)-\bar{\pi}_n^{[J]}(\hat{\alpha}_J),\right.\\
    & \left.  W_i \hat{g}_{\hat{\theta}_1}^{[1]}(X_i, Z_i, W_i; \hat{\beta}_1)-\bar{g}^{w[1]}_{\hat{\theta}_1}(\hat{\beta}_1), \ldots,  W_i \hat{g}_{\hat{\theta}_1}^{[K]}(X_i, Z_i, W_i; \hat{\beta}_K)-\bar{g}^{w[K]}_{\hat{\theta}_1} (\hat{\beta}^K) \right)^{\top},
\end{align*}
$\hat{\alpha}=(\hat{\alpha}_1, \ldots, \hat{\alpha}_J)^{\top}$, $\hat{\beta}=(\hat{\beta}_1, \ldots, \hat{\beta}_K)^{\top}$, and the Lagrange multipliers $\hat{\rho}$ satisfy
\begin{align}
    m^{-1} \sum_{i=1}^{m} \frac{ \hat{h}_i(\hat{\alpha}, \hat{\beta},\hat{\theta}_1) }{ 1 + \hat{\rho}^{\top} \hat{h}_i(\hat{\alpha}, \hat{\beta},\hat{\theta}_1) }=0. \label{nonchange_ELweight}
\end{align}
Next, we consider empirical weights by using the information that $\pi^{[1]}$ is correct, as considered in  \citet{han14}: $\max_{q_1,\ldots, q_{m}} \prod_{i=1}^{m} q_i$, subject to
\begin{align*}
&\sum_{i=1}^{m} q^{(1)}_i=1, \nonumber\\
&\sum_{i=1}^{m} q^{(1)}_i \llp\hat{\pi}^{[j]}(X_i, Z_i, W_i)-\bar{\pi}_n^{[j]}\rrp/\hat{\pi}^{[1]}(X_i, Z_i, W_i)=0,\quad (j=1,\dots, J)\\
&\sum_{i=1}^{m} q^{(1)}_i \llp W_i \hat{g}_{\theta}^{[k]}(X_i, Z_i, W_i)-\bar{g}^{w[k]}_{\theta}\rrp/\hat{\pi}^{[1]}(X_i, Z_i, W_i)=0.\quad (k=1,\dots, K)
\end{align*}
By using the method of Lagrange multipliers again, we obtain
\begin{align*}
    \hat{q}_i= m^{-1}  \frac{1}{ 1 + \hat{\lambda}^{\top} \hat{h}_i(\hat{\alpha}, \hat{\beta},\hat{\theta})/\hat{\pi}^{[1]}(X_i, Z_i, W_i) },\quad (i=1,\dots, m)
\end{align*}
where the Lagrange multipliers $\hat{\lambda}$ satisfy
\begin{align}
    m^{-1} \sum_{i=1}^{m} \frac{ \hat{h}_i(\hat{\alpha}, \hat{\beta},\hat{\theta})/\hat{\pi}^{[1]}(X_i, Z_i, W_i) }{ 1 + \hat{\lambda}^{\top} \hat{h}_i(\hat{\alpha}, \hat{\beta},\hat{\theta})/\hat{\pi}^{[1]}(X_i, Z_i, W_i) }=0.  \label{lamda_satisfy_eq}
\end{align}
It follows from the equation
\begin{align*}
    m^{-1} \sum_{i=1}^{m} \frac{ \hat{h}_i/\hat{\pi}_i^{[1]} }{ 1 + {\lambda}^{\top} \hat{h}_i/\hat{\pi}_i^{[1]} }=(\bar{\pi}_n^{[1]} m)^{-1} \sum_{i=1}^{m} \frac{ \hat{h}_i }{ 1 +  \lp \lambda_1 + 1, \lambda_2, \ldots, \lambda_{J + K}  \rp \hat{h}_i/\bar{\pi}_n^{[1]} },
\end{align*}
that the Lagrange multipliers $\hat{\rho}$ are  $\hat{\rho}_1=(\hat{\lambda}_1 +1)/\bar{\pi}_n^{[1]}$ and $\hat{\rho}_l=\hat{\lambda}_l/\bar{\pi}_n^{[1]},\ l=2,\ldots,J+K$.
Thus, we have
\begin{align*}
    \hat{p}_i^{(1)} = m^{-1} \frac{ \bar{\pi}_n^{[1]}/\hat{\pi}^{[1]}(X_i, Z_i, W_i) }{ 1 + \hat{\lambda}^{\top} \hat{h}_i/\hat{\pi}^{[1]}(X_i, Z_i, W_i) },
\end{align*}
and each $\hat{\theta}_1$ and $\hat{\lambda}$ converges to $\theta^*$ and $0$ in probability, respectively, from the standard theory of empirical-likelihood method in \citet*{qin1994empirical}.

In the second step, the empirical weights are written as
\begin{align*}
    \hat{p}_i^{(2)} = N^{-1} \frac{1}{ 1 + \hat{\upsilon}^{\top} \lp 1-\delta_iW_i\rp \mathcal{C}_{\hat{\theta}_2 }(X_i; \hat{\gamma}) },
\end{align*}
where $\mathcal{C}_{\hat{\theta}_2 }(X_i; \hat{\gamma})=( \hat{C}^{[1]}_{ \hat{\theta}_2 }(X_i; \hat{\gamma}^{[1]})^{\top}, \ldots, \hat{C}^{[L]}_{ \hat{\theta}_2 }(X_i; \hat{\gamma}^{[L]})^{\top} )^{\top}$, $\hat{\gamma}=(  \hat{\gamma}^{[1]}  , \ldots, { \hat{\gamma}^{[L]} } )^{\top}$, and the Lagrange multipliers $\hat{\upsilon}$ satisfy
\begin{align}
    \sum_{i=1}^{N} \frac{  \lp 1-\delta_iW_i\rp \mathcal{C}_{\hat{\theta}_2 }(X_i; \hat{\gamma})  }{  1 + \hat{\upsilon}^{\top} \lp 1-\delta_iW_i\rp \mathcal{C}_{\hat{\theta}_2 }(X_i; \hat{\gamma})  }=0. \label{setting1_step2_eq}
\end{align}
By using the theory in \citet*{qin1994empirical} again, we can show that each $\hat{\theta}_2$ and $\hat{\upsilon}$ converges to $\theta^*$ and $0$ in probability.
Therefore, the two empirical weights and the uniform law of large numbers provide the asymptotic unbiasedness of the estimating equation:
\begin{align*}
    &n\sum_{i=1}^N \hat{p}^{(2)}_i\hat{p}^{(1)}_i\delta_i R_i  W_i  U_{\hat{\theta}_{\EL 1}}(X_i, Y_i)\\
    &=n\frac{\bar{\pi}_n^{[1]}}{m} N^{-1}  \sum_{i=1}^{N} \frac{1}{ 1 + \hat{\upsilon}^{\top} \lp 1-\delta_iW_i\rp \mathcal{C}_{\hat{\theta}_2 }(X_i; \hat{\gamma})^{\top} } \frac{ R_i/\hat{\pi}^{[1]}(X_i, Z_i, W_i) }{ 1 + \hat{\lambda}^{\top} \hat{h}_i/\hat{\pi}^{[1]}(X_i, Z_i, W_i) }\delta_i   W_i  U_{\hat{\theta}_{\EL 1}}(X_i, Y_i)\\
    &=N^{-1} \sum_{i=1}^{N} \frac{R_i}{\hat{\pi}^{[1]}(X_i, Z_i, W_i)} \delta_i   W_i  U_{\hat{\theta}_{\EL 1}}(X_i, Y_i) + o_p(1)\\
    &\xrightarrow{p} E\llp \frac{R}{\pi^{[1]}(X,Z,W)} \delta W U_{\theta^*}(X,Y) \rrp=0.
\end{align*}
Next, we consider when the outcome models include the correct model and suppose that $g^{[1]}$ is the correct model. It follows from the constraint 
\begin{align}
    \sum_{i=1}^{N} \hat{p}_i^{(1)}R_i W_i g^{[1]}(X_i, Z_i, W_i)= n^{-1} \sum_{i=1}^{n} W_ig^{[1]}(X_i, Z_i, W_i) \label{constraint_reg_correct}
\end{align}
that the estimating equation is asymptotically unbiased:
\begin{align*}
    &N\sum_{i=1}^N \hat{p}^{(2)}_i\hat{p}^{(1)}_i\delta_i R_i  W_i  U_{\hat{\theta}_{\EL 1}}(X_i, Y_i)\\
     &=\sum_{i=1}^{N}\hat{p}^{(1)}_i R_i \llp \delta_i W_i  U_{\hat{\theta}_{\EL 1}}(X_i, Y_i) - W_i g^{[1]}(X_i, Z_i, W_i)\rrp + n^{-1} \sum_{i=1}^{n} W_ig^{[1]}(X_i, Z_i, W_i) +o_p(1)\\
     &= m^{-1}\sum_{i=1}^{N} \frac{1}{1 + \hat{\rho}^{\top}\hat{h}_i } R_i\delta_i W_i \llp  U_{\hat{\theta}_{\EL 1}}(X_i, Y_i) - g^{[1]}(X_i, Z_i, W_i)\rrp  +o_p(1)\\
     &\xrightarrow{p} \frac{1}{P(\delta=R=1)} E\lllp \frac{ R\delta W \llp   U_{\theta^*}(X, Y) - g^{[1]}(X, Z, W)\rrp }{ 1 + {\rho^*}^{\top} h^{\dagger}(X,Z,W) } \rrrp=0,
\end{align*}
where 
\begin{align*}
h^{\dagger}(X,Z,W)^{\top}&=( \pi^{[1]}(\alpha_1^*) - \Gamma_1, \ldots, \pi^{[J]}(\alpha_J^*) - \Gamma_J, \\
& \quad \quad W g_{\theta_*}^{[1]}(X, Z, W;\beta_1^*) - \Gamma_{J+1}, \ldots , W_i g_{\theta_*}^{[K]}(X, Z, W;\beta_K^*) - \Gamma_{J+K} ), 
\end{align*}
and $\Gamma_j\,(j=1,\ldots,J)$ and $\Gamma_{J+k}\,(k=1,\ldots,K)$ are the probability limit of $\bar{\pi}_n^{[j]}\, (j=1,\ldots,J)$ and $\bar{g}^{w[k]}_{\hat{\theta}_1}\,(k=1,\ldots,K)$.
Therefore, the estimator $\hat{\theta}_{\EL 1}$ has consistency when one of the $J+K$ models for the response mechanism and the regression function is correct.

We prove the efficiency of $\hat{\theta}_{\EL 1}$.
The Taylor expansion of the left-hand side of 
\eqref{setting1_step2_eq}  around $(\upsilon^{\top}, \gamma^{\top}, \theta^{\top} )=(0^{\top}, {\gamma^*}^{\top}, {\theta^*}^{\top} )$ is
\begin{align*}
    0&= N^{-1/2} \sum_{i=1}^{N} ( 1 - \delta_i W_i) \mathcal{C}_{{\theta}^* }(X_i; {\gamma}^*) \\
    & \quad -N^{-1}\sum_{i=1}^{N} (1 - \delta_i W_i)^2 \mathcal{C}_{\theta^*}(X_i; \gamma^*)^{\otimes 2} N^{1/2}  \hat{\upsilon} +o_p(1).
\end{align*}
With this equation, $\hat{\upsilon}$ can be expanded as
\begin{align}
    N^{1/2}  \hat{\upsilon}  =E\llp (W-1) \mathcal{C}_{\theta^*}(X; \gamma^*)^{\otimes 2}  \rrp^{-1} N^{-1/2} \sum_{i=1}^{N} ( 1 - \delta_i W_i) \mathcal{C}_{\theta^*}(X_i ; \gamma^*) +o_p(1). \label{expansion_upsilon}
\end{align}
In a similar way, the left-hand side of \eqref{lamda_satisfy_eq} can be expanded as
\begin{align}
    0
    &=N^{-1/2} \sum_{i=1}^{N} \delta_i\frac{R_i- \pi_i^{[1]}}{\pi_i^{[1]}} h^{\dagger}(X_i,Z_i,W_i)-
     E
     \biggl\{
     \frac{h^{\dagger}(X,Z,W)^{\otimes 2}}{\pi^{[1]} W} 
     \biggr\}
     N^{1/2} \hat{\lambda}\nonumber\\
     & \quad +N^{-1}\sum_{i=1}^{N} \delta_iR_i
     \biggl\{
     \frac{1}{\lp\pi_i^{[1]}\rp^2} 
     \biggl(
     \frac{\partial \hat{h}_i}{\partial \alpha_1^{\top}} \pi_i^{[1]} - \hat{h}_i\frac{\partial \pi_i^{[1]} }{ \partial \alpha_1 }
     \biggr)
     - \frac{\hat{h}_i}{ \pi_i^{[1]} } 1^{\top} 
     \biggr\}
     N^{1/2} (\hat{\alpha}_1 - \alpha^*)+o_p(1).\label{expansion_lambda}
\end{align}
Recall that our empirical-likelihood estimator is the solution to
\begin{align*}
    Q(\hat{\upsilon}, \hat{\lambda}, \hat{\alpha}, \hat{\beta}, \hat{\gamma}, \hat{\theta}_1, \hat{\theta}_2, \hat{\theta}_{\EL 1})= \sum_{i=1}^{N} \hat{p}_i^{(2)}(\hat{\upsilon}, \hat{\gamma}, \hat{\theta}_2)\hat{p}_i^{(1)}(\hat{\lambda}, \hat{\alpha}, \hat{\beta}, \hat{\theta}_1) \delta_i R_i W_i U_{ \hat{\theta}_{\EL 1} }(X_i,Y_i)=0.
\end{align*}
By substituting $p_i^{(1)}$ and $p_i^{(2)}$ into the above equation and expanding, we have 
\begin{align}
    0&=N^{-1/2} \sum_{i=1}^{N} \lp 1 - \frac{ \hat{\lambda}^{\top} \hat{h}_i/\hat{\pi}_i^{[1]} }{ 1 + \hat{\lambda}^{\top} \hat{h}_i/\hat{\pi}_i^{[1]} } \rp 
    \biggl\{
    1 - \frac{ \hat{\upsilon}^{\top} \lp 1-\delta_iW_i\rp \mathcal{C}_{\hat{\theta}_2 }(X_i; \hat{\gamma}) }{ 1 + \hat{\upsilon}^{\top} \lp 1-\delta_iW_i\rp \mathcal{C}_{\hat{\theta}_2 }(X_i; \hat{\gamma}) } 
    \biggl\}
    \delta_i W_i \frac{R_i}{ \hat{\pi}_i^{[1]} } U_{ \hat{\theta}_{\EL 1} }(X_i,Y_i)\nonumber \\
    &=N^{-1/2} \sum_{i=1}^{N} \frac{R_i}{\pi_i^{[1]}} \delta_iW_i U_{\theta^*}(X_i,Y_i) -E \llp \frac{ U_{\theta^*}(X,Y) }{ \pi^{[1]} } h^{\dagger}(X_i, Z_i, W_i)^{\top}\rrp N^{1/2} \hat{\lambda}\nonumber\\
    & \quad + E \llp (W-1) U_{\theta^*}(X, Y)  \mathcal{C}_{\theta^*}(X;\gamma^*)^{\top}  \rrp N^{1/2} \hat{\upsilon}\nonumber\\
    & \quad + E \llp \frac{U_{\theta^*}(X, Y)}{\pi^{[1]}} \lp \frac{\partial \pi^{[1]}(\alpha_1^*)}{\partial \alpha_1}\rp^{\top} \rrp N^{1/2} \lp \hat{\alpha}_1^* - \alpha_1^* \rp\nonumber\\
    & \quad 
    + E
    \biggl\{
    \frac{\partial U_{\theta^*}(X, Y)}{\partial \theta}
    \biggl\}
    N^{1/2} \lp \hat{\theta}_{\EL 1} -\theta^* \rp +o_p(1).\label{expansion_Q}
\end{align}
It remains to show that terms in \eqref{expansion_Q} reduce to 
\begin{align}
& E \llp (W-1) U_{\theta^*}(X, Y)  \mathcal{C}_{\theta^*}(X;\gamma^*)^{\top}  \rrp N^{1/2} \hat{\upsilon} \nonumber\\
&= E \llp (W-1) C_{\theta^*}^{[1]}  \mathcal{C}_{\theta^*}^{\top}  \rrp E\llp (W-1) \mathcal{C}_{\theta^*}^{\otimes 2}  \rrp^{-1} N^{-1/2} \sum_{i=1}^{N} ( 1 - \delta_i W_i) \mathcal{C}_{\theta^*}(X_i ; \gamma^*) +o_p(1) \nonumber\\
&= N^{-1/2} \sum_{i=1}^{N} (1-\delta_iW_i) C_{\theta^*}^{[1]}(X_i) + o_p(1), \label{thm3_eq1}
\end{align}
and
\begin{align}
    &E \llp \frac{ U_{\theta^*}(X,Y) }{ \pi^{[1]} } h^{\dagger}(X_i, Z_i, W_i)^{\top} \rrp N^{1/2} \hat{\lambda}\nonumber\\
    &=E 
    \biggl(
    \frac{ W g^{[1]} }{\pi^{[1]} W } {h^{\dagger}}^{\top} 
    \biggr)
    E\lp  \frac{{h^{\dagger}}^{\otimes 2}}{\pi^{[1]} W} \rp^{-1} 
    \Biggl[
    N^{-1/2} \sum_{i=1}^{N} \delta_i\frac{R_i- \pi_i^{[1]}}{\pi_i^{[1]}} h_i^{\dagger} 
    \nonumber\\
    & \quad 
    +N^{-1}\sum_{i=1}^{N} \delta_iR_i
    \biggl\{
    \frac{1}{\lp\pi_i^{[1]}\rp^2} 
    \biggl(
    \frac{\partial \hat{h}_i}{\partial \alpha_1^{\top}} \pi_i^{[1]} - \hat{h}_i\frac{\partial \pi_i^{[1]} }{ \partial \alpha_1 }
    \biggr)    - \frac{\hat{h}_i}{ \pi_i^{[1]} } 1^{\top} 
    \biggr\}
    N^{1/2} (\hat{\alpha}_1 - \alpha^*) 
    \Biggr]
    +o_p(1)\nonumber\\
    &= \lp 0,\ldots,0,1,0,\ldots,0 \rp
    \Biggr[
    N^{-1/2} \sum_{i=1}^{N} \delta_i\frac{R_i- \pi_i^{[1]}}{\pi_i^{[1]}} h_i^{\dagger} 
    \nonumber\\
    & \quad 
    +N^{-1}\sum_{i=1}^{N} \delta_iR_i
    \Biggr\{
    \frac{1}{\lp\pi_i^{[1]}\rp^2} 
    \biggl(
    \frac{\partial \hat{h}_i}{\partial \alpha_1^{\top}} \pi_i^{[1]} - \hat{h}_i\frac{\partial \pi_i^{[1]} }{ \partial \alpha_1 }
    \biggr)
    - \frac{\hat{h}_i}{ \pi_i^{[1]} } 1^{\top} 
    \biggr\}
    N^{1/2} (\hat{\alpha}_1 - \alpha^*) 
    \Biggr]
    +o_p(1)\nonumber\\
    &=N^{-1/2} \sum_{i=1}^{N} \delta_i\frac{R_i- \pi_i^{[1]}}{\pi_i^{[1]}} g_i^{[1]} 
    + E \bigg\{ \frac{U_{\theta^*}}{\pi^{[1]}} \lp \frac{\partial \pi^{[1]}(\alpha_1^*)}{\partial \alpha_1}\rp^{\top} \bigg\} N^{1/2} (\hat{\alpha}_1 - \alpha^*)+o_p(1). \label{thm3_eq2}
\end{align}
Then, equations \eqref{expansion_Q}--\eqref{thm3_eq2} reveal that the influence function of the two-step empirical-likelihood estimator in Setting 1 is asymptotically the same as $S_{\eff, 1}$.

We prove the property of $\hat{\theta}_{\EL 2}$.
In setting 2, by using the method of Lagrange multipliers, the second-step empirical weights are 
\begin{align*}
    \hat{p}_i^{(2)}=n^{-1} \frac{1}{ 1 + \hat{\zeta}(W_i^{-1} - \hat{V} )  },\quad (i=1,\dots, n)
\end{align*}
where $\hat{\zeta}$ and $\hat{V}$ satisfy
\begin{align}
    \sum_{i=1}^{n} \frac{\zeta (1/W_i - V)}{ 1 + \zeta (1/W_i - V) }=0, \quad \sum_{i=1}^{n} \frac{\zeta }{ 1 + \zeta (1/W_i - V) } - \frac{N-n}{1-V}=0. \label{cal_of_zeta}
\end{align}
It follows from \eqref{cal_of_zeta} that $\hat{V}= 1 + \hat{\zeta}^{-1} (1 - N/n)$.

First, we prove the consistency of $\hat{\zeta}$ and $\hat{V}$.
By using the same arguments of \citet{qin2002estimation}, after profiling out $p_i^{(2)}$, the log-likelihood is 
$l_1(\xi, V) + l_2(V)$, where $l_1(\zeta, V)= -\sum_{i=1}^{n} \log \{ 1 + \xi  V( 1 - VW_i ) \}$, $l_2(V)= n\log V W_i + (N-n) \log (1-V)$,
and $\xi=\zeta - V^{-1}$ satisfy 
\begin{align*}
    \sum_{i=1}^{n}  \frac{  V\lp 1 - VW_i \rp}{  1 + \xi  V\lp 1 - VW_i \rp  }=0.
\end{align*}
By a similar argument to \citet{qin2002estimation}, if $V$ is in the set $\{ V : \|V\|=N^{-1/3} \}$, we can show that 
$l_1(\xi(V), V)> l_1(\xi(V^*), V^*)$ a.s. and 
$l_1(\xi(V), V) +l_2(W)> l_1(\xi(V^*), V^*)+l_2(W_0)$ a.s.
Therefore, we obtain consistency of $\hat{\zeta}\to P(\delta=1)^{-1}$ and $\hat{V}\to P(\delta=1)$.

It follows analogously from 
\begin{align*}
    N \sum_{i=1}^{N} \hat{p}_i^{(2)}(\hat{\zeta})\hat{p}_i^{(1)}(\hat{\lambda}, \hat{\alpha}, \hat{\beta}, \hat{\theta}_1) \delta_i R_i  U_{ \hat{\theta}_{\EL 2} }(X_i,Y_i)
    = \sum_{i=1}^{N} \hat{p}_i^{(1)}(\hat{\lambda}, \hat{\alpha}, \hat{\beta}, \hat{\theta}_1) \delta_i R_i W_i U_{ \hat{\theta}_{\EL 2} }(X_i,Y_i),
\end{align*}
that our empirical-likelihood-based estimator in Setting 2 has multiple robustness.

By using the Taylor expansion, we obtain
\begin{align}
    N^{1/2} \llp \hat{\zeta} - P(\delta=1)^{-1} \rrp= \frac{1}{P(\delta=1)E(W-1)} N^{-1/2}\sum_{i=1}^{N} (1-\delta_iW_i) +o_p(1). \label{expansion_nu}
\end{align}

Recall that our empirical-likelihood estimator is the solution to
\begin{align*}
    Q(\hat{\zeta}, \hat{\lambda}, \hat{\alpha}, \hat{\beta}, \hat{\theta}_1 , \hat{\theta}_{\EL 2})= \sum_{i=1}^{N} \hat{p}_i^{(2)}(\hat{\zeta})\hat{p}_i^{(1)}(\hat{\lambda}, \hat{\alpha}, \hat{\beta}, \hat{\theta}_1) \delta_i R_i  U_{ \hat{\theta}_{\EL 2} }(X_i,Y_i)=0.
\end{align*}
The Taylor expansion of the estimating equation is
\begin{align}
    0
    &=N^{-1/2} \sum_{i=1}^{N} \delta_iW_i \llp \frac{R_i}{\pi_i^{[1]}}  U_{\theta^*}(X_i,Y_i) +\frac{ \pi_i^{[1]} - R_i}{\pi_i^{[1]}} g_{\theta^*}^{[1]}(X_i, Y_i, W_i;\beta_1^*) \rrp\nonumber\\
    & \quad +P(\delta=1)E\llp (W-1) U_{\theta^*}(X, Y) \rrp N^{1/2} \llp \hat{\zeta} - P(\delta=1)^{-1} \rrp\nonumber\\
    & \quad + E\lp \frac{\partial U_{\theta^*}(X, Y)}{\partial \theta}\rp N^{1/2} \lp \hat{\theta}_{\EL 2} -\theta^* \rp +o_p(1)\label{expansion_Q_set2}.
\end{align}

By substituting \eqref{expansion_nu} into the second term of \eqref{expansion_Q_set2}, we have
\begin{align*}
    &  -E\lp \frac{\partial S_{\eff,2}}{\partial \theta^{\top}}\rp N^{1/2} \lp \hat{\theta}_{\EL 2} -\theta^* \rp\\
    &=N^{-1/2} \sum_{i=1}^{N} \delta_iW_i \llp \frac{R_i}{\pi_i^{[1]}}  U_{\theta^*}(X_i,Y_i) +\frac{ \pi_i^{[1]} - R_i}{\pi_i^{[1]}} g_{\theta^*}^{[1]}(X_i, Y_i, W_i;\beta_1^*) \rrp\\
    & \quad +N^{-1/2} \sum_{i=1}^{N} (1-\delta_iW_i) \frac{E\llp (W-1) U_{\theta^*}(X, Y) \rrp}{E(W-1)} + o_p(1).
\end{align*}
Thus, the influence function of the two-step empirical-likelihood estimator in Setting 2 is asymptotically the same as $S_{\eff, 1}$.

\vspace{2ex}

\noindent 
Proof of Theorem 5.
Because the proof of multiple robustness is almost the same, we prove only the semiparametric efficiency of our proposed estimator in Setting 3.
In Setting 3, the empirical weights are represented by
\begin{align*}
    \hat{p}_i^{(2)}(\nu, \zeta , \tau)= n^{-1}  \frac{1}{ N/n + \nu (W_i^{-1} - 1) +\zeta D^*_\tau(X_i, Z_i) }.
\end{align*}

Recall that our empirical-likelihood estimator is the solution to
\begin{align*}
    Q(\hat{\nu}, \hat{\zeta}, \hat{\tau}, \hat{\lambda}, \hat{\alpha}, \hat{\beta}, \hat{\theta}_1 , \hat{\theta}_{\EL 3})= \sum_{i=1}^{N} \hat{p}_i^{(2)}(\hat{\nu}, \hat{\zeta} , \hat{\tau})\hat{p}_i^{(1)}(\hat{\lambda}, \hat{\alpha}, \hat{\beta}, \hat{\theta}_1) \delta_i R_i  U_{ \hat{\theta}_{\EL 3} }(X_i,Y_i)=0.
\end{align*}
The Taylor expansion of the estimating equation is
\begin{align}
    0
    &=N^{-1/2} \sum_{i=1}^{N} \delta_iW_i \llp \frac{R_i}{\pi_i^{[1]}}  U_{\theta^*}(X_i,Y_i) +\frac{ \pi_i^{[1]} - R_i}{\pi_i^{[1]}} g_{\theta^*}^{[1]}(X_i, Y_i, W_i;\beta_1^*) \rrp\nonumber\\
    & \quad +P(\delta=1)E\llp (W-1) U_{\theta^*}(X, Y) \rrp N^{1/2} \llp \hat{\nu} - P(\delta=1)^{-1} \rrp\nonumber\\
    & \quad -P(\delta=1)E\llp W U_{\theta^*}(X, Y) D^*_\tau(X, Z)^{\top} \rrp N^{1/2} \hat{\zeta}\nonumber\\
    & \quad + E
    \biggl\{
    \frac{\partial U_{\theta^*}(X, Y)}{\partial \theta}
    \biggr\}
    N^{1/2} \lp \hat{\theta}_{\EL 3} -\theta^* \rp +o_p(1)\label{expansion_Q_set3}.
\end{align}
By using the method of Lagrange multipliers and the Taylor expansion, we obtain
\begin{align*}
N^{1/2}\begin{pmatrix} \hat{\nu} - P(\delta=1)^{-1}  \\ \hat{\zeta} \\ \hat{\tau} -  \tau^*\end{pmatrix}=-K^{-1}
\begin{pmatrix} N^{-1/2} \sum_{i=1}^{N} (1 - \delta_iW_i) \\ N^{-1/2} \sum_{i=1}^{N} \delta_iW_i D^*_{\tau^*}(X_i, Z_i)\\ \{P(\delta=1)\}^{-1}\rho \Sigma_1^{-1} N^{1/2} \lp \tilde{\tau} - \tau^* \rp \end{pmatrix} +o_p(1),
\end{align*}{}
where
\begin{align}
K= {\begin{pmatrix}
P(\delta=1) E(1-W) & P(\delta=1) E\{ (W-1) D_{\tau}^* \}^{\top}  & O \\
P(\delta=1) E\{ (W-1) D_{\tau}^* \} & -P(\delta=1) E(
 W{D_{\tau}^*}^{\otimes 2}) & E(\partial D_{\tau}^*/\partial \tau) \\
O & E( \partial D_{\tau}^*/\partial \tau )^{\top}&  P(\delta=1)^{-1} \rho \Sigma_1^{-1} \end{pmatrix}} \label{complex_calculation}.
\end{align}
The terms in \eqref{expansion_Q_set3} can be simplified to
\begin{align*}
    &  - \lp P(\delta=1)E\llp (W-1) U_{\theta^*}\rrp \quad -P(\delta=1)E\llp W U_{\theta^*}(X, Y) {D_{\tau}^*}^{\top} \rrp \quad O \rp K^{-1}\\
    &=(G_1 \quad G_2 \quad G_3),
\end{align*}
where
\begin{align*}
    G_1&= \frac{E\llp (W-1)U_{\theta^*} \rrp}{E(W-1)} - E\lp \frac{\partial U_{\theta^*}}{\partial \theta^{\top}} \rp M E\lp \frac{\partial D_{\tau}^*}{ \partial \tau^{\top}} \rp^{-1} \frac{E\llp (W-1)D_{\tau}^* \rrp}{ E(W-1) },\\
    G_2&= - E\lp \frac{\partial U_{\theta^*}}{ \partial \theta^{\top} } \rp M E\lp \frac{\partial D_{\tau}^*}{ \partial \tau^{\top}} \rp^{-1},\\
     G_3&=P(\delta=1) E\lp \frac{\partial U_{\theta^*}}{ \partial \theta^{\top} } \rp M \frac{\Sigma_1}{\rho},\\
     M&= E\lp \phi_{\eff} \eta_{\eff}^{\top} \rp \llp \frac{\Sigma_1}{\rho} +E\lp \eta_{\eff}^{\otimes2} \rp  \rrp^{-1},
\end{align*}
and $\phi_{\eff}$ and $\eta_{\eff}$ are the efficient influence functions for $\theta$ and $\tau$ based on the internal individual data.
By the Taylor expansion \eqref{expansion_Q_set3} and \eqref{complex_calculation}, we can show that
\begin{align*}
    &  -N^{1/2} \lp \hat{\theta}_{\EL 3} -\theta^* \rp \\
    &=E\lp \frac{\partial U_{\theta^*}}{ \partial \theta^{\top} } \rp^{-1}N^{-1/2} \sum_{i=1}^{N} \delta_iW_i 
    \biggl(
    \frac{R_i}{\pi_i^{[1]}}  U_{\theta^*} +\frac{ \pi_i^{[1]} - R_i}{\pi_i^{[1]}} g_{\theta^*}^{[1]} 
    \biggr)
    \nonumber\\
    & \quad +E\lp \frac{\partial U_{\theta^*}}{ \partial \theta^{\top} } \rp^{-1} G_1 N^{-1/2} \sum_{i=1}^{N} (1 - \delta_iW_i) + E\lp \frac{\partial U_{\theta^*}}{ \partial \theta^{\top} } \rp^{-1} G_2 N^{-1/2} \sum_{i=1}^{N} \delta_iW_i D_{\tau}^*(X_i, Z_i) \nonumber\\
    & \quad + E\lp \frac{\partial U_{\theta^*}}{ \partial \theta^{\top} } \rp^{-1} G_3 \frac{1}{P(\delta=1)} \rho \Sigma_1^{-1} N^{1/2} \lp \tilde{\tau} - \tau^* \rp +o_p(1)\\
    &=E\lp \frac{\partial U_{\theta^*}}{ \partial \theta^{\top} } \rp^{-1}N^{-1/2} \sum_{i=1}^{N} 
    \biggl\{
    \delta_iW_i 
    \biggl(
    \frac{R_i}{\pi_i^{[1]}}  U_{\theta^*} +\frac{ \pi_i^{[1]} - R_i}{\pi_i^{[1]}} g_{\theta^*}^{[1]} 
    \biggr)
    + (1 - \delta_iW_i)\frac{E\llp (W-1)U_{\theta^*} \rrp}{E(W-1)}  
    \biggr\}
    \nonumber\\
    & \quad - M E\lp \frac{\partial D_{\tau}^*}{ \partial \tau^{\top}} \rp^{-1} N^{-1/2} \sum_{i=1}^{N} (1 - \delta_iW_i) \frac{E\llp (W-1)D_{\tau}^* \rrp}{ E(W-1) } \\
    & \quad -  M E\lp \frac{\partial D_{\tau}^*}{ \partial \tau^{\top}} \rp^{-1} N^{-1/2} \sum_{i=1}^{N} \delta_iW_i D_{\tau}^*(X_i, Z_i) + MN^{1/2} \lp \tilde{\tau} - \tau^* \rp +o_p(1)\\
    &= N^{-1/2} \sum_{i=1}^{N} \lp \phi_{\eff, i} - M \eta_{\eff, i} \rp + MN^{1/2} \lp \tilde{\tau} - \tau^* \rp +o_p(1).
\end{align*}
Thus, the influence function of the two-step empirical-likelihood estimator in Setting 3 is asymptotically the same.

\bibliographystyle{unsrtnat}
\bibliography{ref2}  

@string{jasa = "J. Am. Statist. Assoc."}

@string{annstat = "Ann. Statist."}

@string{smj = "Survey Methodol."}

@string{isr="Int. Statist. Rev."}

@string{cjs="Can. J. Statist."}

@string{economj = "Econom. J."}

@string{statsin = "Statist. Sin."}

@book{bickel1998,
   author = {Bickel, P.~J. and  Klaassen, C.~A.~J. and Ritov, Y.}, 
   year = {1998}, 
   title = {Efficient and Adaptive Estimation for Semiparametric Models}, publisher = {Springer-Verlag}, 
   address = {New York}
}

@article{Binder1983,
 author = {Binder, D. A.},
 year =  {1983},
 title = {On the variances of asymptotically normal estimators from complex surveys},
 journal = isr,
 volume = { 51},
 pages = {279--292}
 }

@book{Chambers2003,
  title={Analysis of Survey Data},
  author={ Chambers, R.L. and Skinner, C. J. (Eds.) },
  year={2003},
  publisher={ John Wiley $\&$ Sons}
}

@article{chernozhukov18,
	Author = {Chernozhukov, Victor and Chetverikov, Denis and Demirer, Mert and Duflo, Esther and Hansen, Christian and Newey, Whitney and Robins, James},
	Journal = economj,
	Number = {1},
	Pages = {C1--C68},
	Title = {Double/debiased machine learning for treatment and structural parameters},
	Volume = {21},
	Year = {2018}}

@article{chatterjee2016,
   author = {Chatterjee, N. and  Chen, Y-H and Maas, P. and Carroll, R.J.},
   title = { Constrained maximum likelihood estimation for model calibration using summary-level information from external big data sources}, 
   year = {2016},
   journal = jasa, 
   volume = {111},
   pages = {107--117}
}

@article{chen17_mr,
  title={Multiply robust imputation procedures for the treatment of item nonresponse in surveys},
  author={Chen, S. and Haziza, D.},
  journal={Biometrika},
  volume={104},
  pages={439--453},
  year={2017}
}

@article{chen21,
    title = {Sample empirical likelihood approach under complex survey design with scrambled responses}, 
    author = {Chen, S. and Zhao, Y. and Wang, Y. }, 
    year = {2021},
    journal = smj, 
    volume = {47}, 
    pages = {59--75}
}

@article{duan10,
    title = {Optimal estimation in surrogate outcome regression problems}, 
    author = {Duan, X. and Qin, J. and Wang, Q.}, 
    year = {2010},
    journal = cjs, 
    volume = {38}, 
    pages = {633--646}
}

@book{fuller09,
  title={Sampling Statistics},
  author = {Fuller, W.~A.},
  year = 2009,
  publisher = {Wiley}  
}

@article{godambe09,
  title = {Estimating Functions and Survey Sampling},
  author = {Godambe, V. P. and Thompson, M. E.},
  year = 2009,
  journal = {Handbook of Statistics, vol. 29B. Amsterdam: Elsevier}, 
  pages = {455--487}
}

@inproceedings{hajek71,
   author = {Hajek, J.},
   year =  {1971},
   title = {Comment \textit{on An essay on the logical foundations of survey sampling by Basu, D.}},
   booktitle = {Foundations of Statistical Inference (Godambe, V.P. and Sprott, D.A. eds.)},
   publisher = {Holt, Rinehart and Winston},
   pages = {236}
}

@article{han14,
  title={Multiply Robust Estimation in Regression Analysis With Missing Data},
  author={Han, P.},
  journal=jasa,
  volume={109},
  pages={1159--1173},
  year={2014}
}

@article{HT1952,
    author = {Horvitz, D. G. and Thompson, D. J.},
    title = {A generalization of sampling without replacement from a finite universe},
    journal = jasa,
    volume = {47},
    pages = {663--685},
    year = {1952}
}

@article{hu2022paradoxes,
  title={Paradoxes and resolutions for semiparametric fusion of individual and summary data},
  author={Hu, Wenjie and Wang, Ruoyu and Li, Wei and Miao, Wang},
  journal={arXiv preprint arXiv:2210.00200},
  year={2022}
}

@article{kim14_dr,
	Author = {Kim, J. K. and Haziza, D.},
	Journal = statsin,
	Pages = {375--394},
	Title = {Doubly Robust Inference With Missing Data in Survey Sampling},
	Volume = {24},
	Year = {2014}
}

@article{kim13,
	Author = {Kim, J. K. and Skinner, C. J.},
	Journal = {Biometrika},
	Pages = {385--398},
	Title = {Weighting in survey analysis under informative sampling},
	Volume = {100},
	Year = {2013}
}

@article{kundu19,
  title={Generalized meta-analysis for multiple regression models across studies with disparate covariate information},
  author={Kundu, P. and Tang, R. and Chatterjee, N.},
  journal={Biometrika},
  volume={106},
  pages={567--585},
  year={2019}
}

@article{morikawa22,
   author = {Morikawa, K. and Terada, Y. and Kim, J. K.},
   title = {Semiparametric adaptive estimation under informative sampling}, 
   year = 2025,
   journal = {To appear in Ann. Statist.},
note = {Available the arXiv preprint at arXiv:2208.06039}
}

@article{parker23,
    title = {A Bayesian functional data model for surveys collected under informative sampling with application to mortality estimation using NHANES}, 
    author = {Parker, P. A. and Holan, S. H.}, 
    year = {2023},
    journal = {Biometrics}, 
    volume = {79}, 
    pages = {1397--1408}
}

@article{Pfeffermann1993,
  AUTHOR =       {    Pfeffermann, D.   },
  TITLE =        {The role of sampling weights when modeling survey data },
  JOURNAL =      isr,
  YEAR =         { 1993},
  volume =       { 61},
  number =       {},
  pages =        {317-337 }
}

@article{pfe99,
    author = "D. Pfeffermann and M. Sverchkov",
    title = {Parametric and Semi-parametric Estimation of Regression Models Fitted to Survey Data},
    year = {1999},
    journal = {Sankhy\={a}, Series B},
    volume = {61},
    pages = {166--186}
}

@article{qin09,
  title={Empirical Likelihood in Missing Data Problems},
  author={Qin, J. and Zhang, B. and Leung, H. Y.},
  journal= jasa,
  pages={1492--1503},
  year={2009}
}

@article{robins94,
         author = "J.~M. Robins and A.  Rotnitzky and  L.~P. Zhao",
         title = "Estimation of regression coefficients when some regressors are not always observed",
         journal = jasa,
         year = 1994,
         volume = 89,
         pages = "846-866"
}

@article{rubin1976,
  title={Inference and missing data},
  author={Rubin, Donald B},
  journal={Biometrika},
  volume={63},
  number={3},
  pages={581--592},
  year={1976},
  publisher={Oxford University Press}
}

@book{tsiatis2006,            
   author = {Tsiatis, A.~A.}, 
   title = {Semiparametric Theory and Missing Data}, 
   publisher = {Springer-Verlag}, 
   year = {2006},
   address = {New York}
}

@article{zhang2020generalized,
  title={Generalized integration model for improved statistical inference by leveraging external summary data},
  author={Zhang, Han and Deng, Lu and Schiffman, Mark and Qin, Jing and Yu, Kai},
  journal={Biometrika},
  volume={107},
  number={3},
  pages={689--703},
  year={2020},
  publisher={Oxford University Press}
}

@article{qin1994empirical,
  title={Empirical likelihood and general estimating equations},
  author={Qin, Jin and Lawless, Jerry},
  journal=annstat,
  volume={22},
  number={1},
  pages={300--325},
  year={1994},
  publisher={Institute of Mathematical Statistics}
}

@article{qin2002estimation,
  title={Estimation with survey data under nonignorable nonresponse or informative sampling},
  author={Qin, Jing and Leung, Denis and Shao, Jun},
  journal=jasa,
  volume={97},
  number={457},
  pages={193--200},
  year={2002},
  publisher={Taylor \& Francis}
}

@article{rao2010,
  title={ Bayesian pseudo-empirical-likelihood intervals for complex surveys},
  author={J. N. K. Rao and Changbao Wu},
  journal={J. R. Statist. Soc. B},
  volume={72},
  pages={533--544},
  year={2010}
}

@article{zhao2020,
  title={Bayesian Empirical Likelihood Inference with Complex Survey Data},
  author={Puying Zhao and Malay Ghosh and J. N. K. Rao and Changbao Wu},
  journal={J. R. Statist. Soc. B},
  volume={82},
  pages={155--174},
  year={2020}
}

@article{berg16,
  title={Imputation under informative sampling},
  author={Berg, E. and  Kim, J.K. and Skinner, C.J.},
  journal={J. Surv. Stat. Methodol.},
  volume={4},
  pages={436--462},
  year={2016}
}

@article{pferffermann09,
   author = {Pfeffermann, D. and Sverchkov, M.},
   year = {2009},
   title = {Inference under informative sampling},
   journal = {Handbook of Statistics, vol. 29B. Amsterdam: Elsevier}, 
  pages = {455--487}
}

@article{little03,
   author = {Little, R.J.A.},
   year = {2003},
   title = {Bayesian methods for unit and item nonresponse},
   journal = {{\rm In Chambers,
R.L. and Skinner, C. J. Analysis of Survey Data. Wiley}, Chichester}, 
  pages = {289--306}
}

@article{rotnitzky1997analysis,
  title={Analysis of semi-parametric regression models with non-ignorable non-response},
  author={Rotnitzky, Andrea and Robins, James},
  journal={Statistics in medicine},
  volume={16},
  number={1},
  pages={81--102},
  year={1997},
  publisher={Wiley Online Library}
}

@article{mcneney2000application,
  title={Application of convolution theorems in semiparametric models with non-iid data},
  author={McNeney, Brad and Wellner, Jon A},
  journal={Journal of Statistical Planning and Inference},
  volume={91},
  number={2},
  pages={441--480},
  year={2000},
  publisher={Elsevier}
}






\end{document}